\documentclass[]{jfm}
\usepackage{graphicx}
\usepackage{lineno}
\usepackage{epstopdf, epsfig}
\usepackage{subcaption}
\usepackage{bm}
\usepackage{amsmath,amsfonts,amssymb}
\usepackage{physics}
\usepackage{graphicx}
\usepackage{bm}
\usepackage{color}
\usepackage{hyperref}
\hypersetup{
    colorlinks=true,
    linkcolor=blue,
    citecolor=blue,
    urlcolor=blue,
    pdftitle={Jet--wall paper}
}
\usepackage{hyperref}
\usepackage[mathscr]{euscript}
\usepackage{todonotes}
\usepackage{caption}
\makeatletter
\@ifundefined{backsection}{\newcommand{\backsection}{}}{}
\renewcommand{\backsection}[2][Acknowledgements]{\par\begingroup\small
  \vskip6pt\noindent\textbf{#1.}\ #2\par\endgroup}
\makeatother
\usepackage[most]{tcolorbox}

\title{Stratified jet}
\shortauthor{H.D. Vu \& K. Deguchi}

\title{Stratified jet with wall effects}
\author{Hai Duc Vu\aff{1,2} and Kengo Deguchi\aff{1}\corresp{\email{kengo.deguchi@monash.edu}}}

\affiliation{
\aff{1}School of Mathematics, Monash University, Victoria 3800, Australia
\aff{2}Department of Artificial Intelligence, Monash University, Victoria 3800, Australia}

\begin{document}
%\linenumbers
\maketitle

\begin{abstract}
%[check the draft carefully, fill the missing info, fix errors etc.][yes I am reading now, will take a day or 2 to verify all the details.]
%[We need statement and GA for submission]
% \Kengo{Some statements I think of for now. Feel free to edit please

% Jets in stratified fluids are ubiquitous in the atmosphere and ocean, and are almost always modelled as unbounded. We show that distant boundaries are not a detail. For a Bickley jet in a uniformly stratified channel, we compute the first exact coherent structures for a stratified jet and identify the first radiating instabilities in uniform stratification: their internal waves carry momentum into the ambient and decelerate the jet. Whether these instabilities survive at high Reynolds number is decided at the walls: no-slip Stokes layers extinguish them; slip walls do not. A further viscous instability persists above the Miles–Howard threshold, with or without walls. [this is not abstract.. but anyway do  not worry about it]
% }
%\textcolor{black}{[.. no comment of this paper in the other paper?][as in citation of this paper in the jet intrusion paper?][yes .. otherwise oceanographers may not read this paper][oh ofc I am planning to put it in the discussion. Going through the jet paper today for finalisation]}
Jets in stratified flows are ubiquitous in geophysics. A fundamental question is whether the instabilities and coherent structures generated by the flow interact with the physical boundaries and, if so, how.
%
%A common belief is that a jet and its surrounding walls do not interact when they are far apart. 
%
\textcolor{black}{To this end, a simple canonical model flow is employed:} a parallel Bickley jet confined between walls in a uniformly stratified fluid.
We first show that short wavelength Kelvin-Helmholtz instability can generate asymmetry in the mean flow. Although the fluctuations decay rapidly away from the jet, \textcolor{black}{the asymmetric mean flow does not. 
%when it is asymmetric about the jet centreline.
}
We then examine long-wavelength coherent structures that can emit internal gravity waves, which decay slowly towards the walls or, in some cases, radiate away from the jet. These waves transport momentum and distort the mean flow, reducing the jet centreline velocity significantly. When radiation occurs, whether the instability persists at high Reynolds numbers depends strongly on the boundary conditions imposed at the walls. 
Moreover, we found an instability and associated nonlinear coherent structures that persist at Richardson numbers above the Miles-Howard criterion. The existence of this viscous mode does not \textcolor{black}{rely} on the presence of the walls.
\end{abstract}

\section{Introduction}

The interaction between stratification and shear is a core problem in fluid mechanics. To understand the underlying physical mechanisms, a myriad of theoretical studies have been carried out using simple canonical model flows. 
As for example reviewed in \citet{drazin1966hydrodynamic} and \citet{peltier2003mixing},
compact shear layers, such as jets and mixing layers, are of particular interest in geophysical studies as they %are believed to 
play a central role in atmospheric and oceanic dynamics.
Representative examples include the Bickley jet \citep{schlichting1933laminare,bickley1937plane} %\textcolor{black}{[add Schlichting 1933]} 
and the Holmboe/Hazel hyperbolic tangent mixing shear layer profile \citep{holmboe1962behavior,hazel1972numerical}.

The use of simple model base flows was essential for making progress before the advent of modern computing. Of particular relevance to the present work is the study of \citet{drazin1966hydrodynamic}, who derived
an analytical expression for the neutral linear stability curve of a uniformly stratified Bickley jet.
%(this work generalised the earlier unstratified results of Lipps (1962)).
As in the majority of theoretical studies, the inviscid assumption was invoked to reduce the governing equations to the Taylor-Goldstein equation \citep{taylor1931effect,goldstein1931stability}.
\citet{hazel1972numerical} solved this equation numerically
and confirmed the predictions of \citet{drazin1966hydrodynamic}.
\citet{sutherland1992stability} later provided a more detailed analysis of the growth rates of the unstable modes and established a link to spatial instability.

As first demonstrated by \citet{miles1961stability}, neutral modes of the Taylor-Goldstein equation may exhibit singular behaviour, and thus traditional computational methods struggle to find accurate neutral curves. Robust numerical techniques for computing such singular solutions have been developed for a wide range of base flows \citep[e.g.][]{rees2014general,hirota2016stability}.
The singularity occurs at the critical level, where the phase speed of the disturbance coincides with the base flow velocity. The appropriate mathematical framework for describing this phenomenon is matched asymptotic expansions, in which the singularity is regularised by viscous and/or nonlinear effects within a thin region known as the critical layer \citep[see][]{miles1961stability,baldwin1970critical,kelly1970nonlinear,brown1981evolution,churilov1988nonlinear,troitskaya1991viscous, maslowe1972generation, maslowe1973finite}.

\citet{drazin1979normal} presented a comprehensive classification of solutions to the Taylor-Goldstein equation in an infinite domain,  one of the key criteria being the presence or absence of a critical level. Furthermore, they distinguished between bounded states, for which the perturbations decay in the far field, and unbounded states, for which they do not. The latter class includes perturbations that oscillate at infinity, which are physically associated with the radiation of internal gravity waves.
\citet{sutherland1994internal} numerically identified such radiating modes in the Bickley jet and the hyperbolic tangent mixing layer by solving the Taylor-Goldstein equation.
It can be readily shown that radiation requires non-vanishing stratification in the far field.
However, since uniform background stratification struggles to produce radiating modes, a suitably varying background density field is adopted instead; the same strategy was used in the nonlinear viscous simulations of \citet{sutherland1994turbulence}.
There is a separate line of research investigating the generation of radiating internal gravity waves using Taylor-Goldstein equation by \citet{dunkerton1997shear} and \citet{kwasniok2003radiating}. Those authors used oscillatory base flows with zero mean shear, together with a buoyancy frequency that oscillates around a constant value. Under uniform stratification, the existence of \textcolor{black}{unstable radiating} modes in a Bickley jet is unknown.
%\Kengo{add unstable here?}

At about the same time as Hazel's seminal work, the first stability analysis to include the effects of diffusivity was carried out by \citet{maslowe1971stability} using the Holmboe hyperbolic tangent profile. Interestingly, subsequent studies showed that stability calculations incorporating finite viscosity exhibit behaviour that is not anticipated by  inviscid theories.
For example,
%Gage (1972),
\citet{miller1972prandtl}, \citet{gage1974linear} %[Linear viscous stability analysis of the stratified Bickley jet] 
and \citet{miller1988viscous} numerically demonstrated, for various smooth and non-smooth base \textcolor{black}{flows,} % in a finite domain
that instability may arise even when the local Richardson number exceeds 1/4 everywhere. Such behaviour is impossible for unstable Taylor-Goldstein solutions because of the well-known Miles-Howard theorem \citep{miles1961stability,howard1961note}. 
%The physical mechanism of those peculiar modes is generally attributed to the overreflection of internal gravity waves at the critical level \citep{booker1967critical,jones1968reflexion}.
More recently, \citet{parker2019kelvin} 
%and 
%\citet{parker2020viscous} 
identified an unstable mode in the strongly stratified regime of Hazel's model profile. 
\textcolor{black}{The phase speed of this mode exceeds the base flow velocity over a wide range of parameters (see \citet{parker2020viscous}),} so that no critical level exists; such behaviour is likewise impossible for inviscidly unstable modes because of Howard's semicircle theorem \citep{howard1961note}. 
\textcolor{black}{\citet{parker2020viscous} already elucidated this seemingly paradoxical numerical result.}
%The seemingly paradoxical numerical results described above were theoretically resolved by.
In the high-Reynolds-number asymptotic analysis,
the leading-order solution is the neutrally stable Taylor-Goldstein solution (which belongs to one of the classes in \citealt{drazin1979normal}), while the small positive growth rate arises from the next order viscous correction.
The neutral inviscid solution lies outside the scope of the Miles-Howard theorem and Howard's semicircle theorem, so there is no contradiction. Instability that emerges at higher order has also been observed in other shear flows \citep[see e.g.][]{duck1994linear,deguchi2025instability}, and are referred to as viscous modes.

Around the beginning of the twenty-first century, direct numerical simulation (DNS) studies of stratified jets became active; a summary of early efforts can be found in \citet{tse2003quasi}.
Most of these studies have been motivated by their geophysical applications. For example, \citet{pham2011mixing} investigated the interaction between a jet and a shear layer, while the recent study of \citet{vu2026planar} on the intrusive gravity current process into a uniformly stratified ambient fluid represents another related example.
Nevertheless, for canonical model flows such as the Bickley jet, relatively few studies have followed the pioneering works of \citet{sutherland1994turbulence} and \citet{sutherland1994internal}. This contrasts sharply with hyperbolic tangent shear layers, for which DNS studies have been regularly reported \citep[see][and references therein]{vandine2021turbulent}.
%[memo: texts below need to be modified, will do]
In DNS the computational domain size needs to be chosen carefully. A domain that is too small restricts the range of vortical structures represented in the simulation, while a larger domain imposes limitations on the achievable Reynolds number and numerical resolution. 
In the vertical direction, the unbounded domain assumption commonly adopted in theoretical studies cannot be employed. 
For stratified shear flows, radiating internal gravity waves can be generated, so inevitably, how the boundary conditions are imposed also becomes important. 
\textcolor{black}{Efforts to at least partially eliminate boundary effects are made by matching the flow field in the bulk to the far-field asymptotic solution or by introducing a near-wall sponge region (see \citet{miller1972prandtl} and  \citet{sutherland1994internal}, for example).}

In many applications, boundaries inevitably exist somewhere in the physical system; for example, atmospheric jets can interact with Earth's surface, \textcolor{black}{while intrusion gravity currents can reach the sea surface. }
A systematic investigation of how far-field boundaries influence instability and/or nonlinear coherent structures induced by jets remains an underexplored research area.  
The interaction between a jet and a wall depends on a number of parameters, including not only the jet-wall distance but also the stratification strength, Reynolds number, and typical wavelength.
Thus, extracting fundamental insights from numerical simulations of realistic flow configurations is challenging.

%, and consequently numerical studies often emphasise the robustness of solutions with respect to changes in the domain size. 

%Our aim is to investigate the wall--jet interaction 

This paper presents the first systematic investigation of a stratified jet over a wide range of parameters, choosing the simplest possible setting: a Bickley jet in a uniformly stratified fluid confined within a channel.
Our interest includes how the linear instability and nonlinear coherent structures interact with the walls. 
We restrict attention to the two-dimensional case under the parallel flow approximation, but even in this setting a detailed parameter study using DNS remains unfeasible. We therefore employ bifurcation analysis based on Newton's method. The solutions found by this approach are what are now referred to as exact coherent structures (ECS) in the wall bounded shear flow community \citep{kerswell2005recent,eckhardt2007turbulence,graham2021exact}. 

ECS satisfy the governing equations exactly (up to numerical accuracy) and are thus particularly amenable to theoretical analysis. On the other hand, they are often unstable, and establishing their relationship to the dynamics requires a detailed analysis based on dynamical systems theory \citep[see e.g.][]{kawahara2012significance,wang2025mathematically}. While there have been studies of wall-bounded stably stratified shear flows in small periodic domains \citep[e.g.][]{clever1992three,deguchi2017scaling,olvera2017exact,langham2020stably}, %clever1997tertiary ! actually no stratification
corresponding studies of compact shear layers remains sparse.
Indeed, the only works of which we are aware are those of \citet{parker2019kelvin, Parker_Caulfield_Kerswell_2021} %\textcolor{black}{[add Parker et al. 2021]} 
and \citet{deguchi2018free}. The latter work studied three-dimensional coherent structures generated in the far field of a spatially developing, unstratified Bickley jet and is therefore not directly related to the present work.

The paper is structured as follows. The next section introduces the mathematical formulation of our setup. To present the numerical results systematically, we begin in section 3 by establishing two baseline results; the first concerns the linear stability, while the second presents a bifurcation analysis of the unstratified Bickley jet. 
The subsequent sections are mainly devoted to investigating the effects of stratification and the jet-wall distance on the linear stability and ECS. Section 4 examines the short-wavelength regime, section 5 the long-wavelength regime, and section 6 the relatively strongly stratified regime. Finally, section 7 discusses the results and concludes the paper. 

%\subsection{Bickley jet, uniform stratification, Taylor-Goldstein}
%Lipps (1962): analytic neutral curve for unstratified Bickley. 
%Howard \& Drazin (1966) generalised this result for stratified case.
%Howard \& Drazin (1966) work is commented in Hazel (1972) ... checked with numerical solution.

%Sutherland \& Peltier (1992): TG numerically solved for Bickley uniform stratification (very important ref) There are comments for Sato 1960; Sato \& Kuriki 1961

%Numerical methods for TG are also presented in Rees \& Monahan (2014), Hirota \& Morrison (2016).

%\subsection{Bickley jet, uniform stratification, viscous linear stability}
%Maslowe Thompson (1971): first attempt to solve viscous problem, tanh profile used. (stability equation derived in Koppel (1964))

%For tanh, PCK19, PCK20 (Hazel)
%Viscous mode .. everywhere phase speed is greater than unity.

%\subsection{variable stratification and radiation}
%Sutherland, Caulfield \& Peltier 1994: Bickley jet, $N^2=J \tanh^2(z/R)$
%TG. They also did nonlinear viscous simulation (sponge layer technique near the boundary). Radiating modes discussed. 

\section{Formulation of the problem}

Consider a jet of a typical speed $u_0$ and width $l_0$, propagating along the centre of a channel. 
For simplicity, we assume that the kinematic viscosity $\nu$, thermal diffusivity $\kappa$, and coefficient of thermal expansion $\gamma$ of the fluid are constant. 
Assuming further that 
the fluid density remains close to the constant reference value $\rho_0$, the evolution of 
the velocity $\mathbf{u}_*$, pressure $p_*$, and the density deviation $\rho_*$ is governed by the Boussinesq equations:
\begin{subequations}
\begin{eqnarray}
\rho_0(\partial_{t_*}+\mathbf{u}_*\cdot \nabla_*)\mathbf{u}_*=-\nabla_* p_*+\rho_0 \nu \nabla_*^2 \mathbf{u}_*-\rho_*g \mathbf{e}_y+F_*\mathbf{e}_x,\\
\nabla_*\cdot \mathbf{u}_*=0,\\
(\partial_{t_*}+\mathbf{u}_*\cdot \nabla_*)\rho_*=\kappa \nabla_*^2 \rho_*.\label{tempdim}
%\\ \rho_*=\rho_0[1-\gamma(\theta_*-\theta_0)].
\end{eqnarray}
\end{subequations}
Here Cartesian coordinates $(x_*,y_*)$ are used, with the $x_*$ axis aligned with the jet direction. 
In the present study, we restrict attention to flows that are independent of the third spatial coordinate. 
Using the unit vectors $\mathbf{e}_x$ and $\mathbf{e}_y$, the velocity can be written in component form as $\mathbf{u}_*=[u_*,v_*]=u_*\mathbf{e}_x+v_*\mathbf{e}_y$. The dimensional gradient operator is denoted by $\nabla_*=[\partial_{x_*},\partial_{y_*}]$. 
 
The channel walls are placed at $y_*=\pm Hl_0$, where temperature is held fixed at $\theta_*=\theta_0\pm H\theta_1$. Recalling the temperature and density deviations are related as $\rho_*=\rho_0(1-\gamma(\theta_*-\theta_0))$ under the Boussinesq approximation, the base density is obtained as $\rho_*=\rho_0(1-\gamma \theta_1 y_*/l_0)$. The base jet is maintained parallel by an external forcing $F_*(y_*)$.

Using the non-dimensional variables $t=(u_0/l_0)t_*$ and $[x,y]=[x_*,y_*]/l_0$, 
we find that
$[u,v]=[u_*,v_*]/u_0$, 
$p=(p_*+\rho_0gy_*)/\rho_0u_0^2$, and $\rho=(\rho_*-\rho_0)/\gamma \rho_0 \theta_1$ satisfy
\begin{subequations}\label{fulleq}
\begin{eqnarray}
(\partial_t+\mathbf{u}\cdot \nabla)\mathbf{u}=-\nabla p+ Re^{-1}\nabla^2 \mathbf{u}- J\rho \mathbf{e}_y+Re^{-1}F\mathbf{e}_x,\label{momentumeq}\\
\nabla\cdot \mathbf{u}=0,\\
(\partial_t+\mathbf{u}\cdot \nabla)\rho=Re^{-1}Pr^{-1} \nabla^2 \rho,\label{densityeq}
\end{eqnarray}
\end{subequations}
where $\nabla=[\partial_{x},\partial_{y}]$. 
The Reynolds number, bulk Richardson number, and Prandtl number are defined as follows:
\begin{eqnarray}
Re=\frac{u_0l_0}{\nu},\qquad 
%Ra=-\frac{g h^3\rho_1}{\nu\kappa\rho_0}
J=\frac{\gamma \theta_1 g l_0}{u_0^2}
,\qquad
Pr=\frac{\nu}{\kappa}.
\end{eqnarray}
The forcing  $F(y)=(l_0^2/\rho_0\nu u_0)F_*(y_*)=(2\text{sech}^2(y)-4\text{tanh}^2(y))\text{sech}^2(y)$ is chosen to support the \textcolor{black}{parallel Bickley jet profile.}
That is, the base flow of the system is given by
\begin{eqnarray}
u=U(y)=\text{sech}^2(y),\qquad v=0,\qquad \rho=-y, \qquad p=\frac{Jy^2}{2}.\label{base}
\end{eqnarray}
In the non-dimensional coordinates, the Bickley jet has a vorticity thickness of  $\approx$1.3 and a 99\% thickness of $\approx$3. 

\textcolor{black}{Unless otherwise noted, we use the no-slip boundary conditions 
\begin{eqnarray}
(u-U)=0,\qquad v=0\qquad \text{at}\qquad y=\pm H \label{noslipBC}
\end{eqnarray}
Here the conditions are imposed for the perturbation velocity relative to the base flow $\tilde{\mathbf{u}}=[u-U,v]$; this is justified as we typically take $H$ to be greater than 10.
%we enforce the perturbation velocity relative to the base flow $\tilde{\mathbf{u}}=[u-U,v]$ to vanish at the walls at $y=\pm H$. This is justified as we typically take $H$ to be greater than 10. 
For some computations the slip (no-stress) boundary conditions 
\begin{eqnarray}
\partial_y(u-U)=0,\qquad v=0\qquad \text{at}\qquad y=\pm H \label{slipBC}
\end{eqnarray}
are employed. For both cases, since the temperature is fixed at the walls, the density perturbation $\tilde{\rho}=\rho+y$ must vanish at $y=\pm H$. }

When the perturbations are small, the governing equations can be linearised. Introducing the streamfunction for the velocity perturbation as $\tilde{u}=\partial_y\tilde{\psi}, \tilde{v}=-\partial_x\tilde{\psi}$, the continuity is satisfied automatically, and the momentum equations can be combined into a single equation.
The normal mode ansatz $\tilde{\psi}=\hat{\psi}(y)e^{ik(x-c t)}$, $\tilde{\rho}=\hat{\rho}(y)e^{ik(x-c t)}$, where $k$ and $c$ are the wavenumber and the phase speed, respectively, yields the well-known stability equations
\begin{subequations}\label{visstab}
\begin{eqnarray}
(U-c)(\hat{\psi}''-k^2\hat{\psi})-U''\hat{\psi}-J\hat{\rho}=\frac{1}{ikRe}(\hat{\psi}''''-2k^2\hat{\psi}''+k^4\hat{\psi}),\label{25eq1}\\
(U-c)\hat{\rho}+\hat{\psi}=\frac{1}{ikRePr}(\hat{\rho}''-k^2\hat{\rho}).
\end{eqnarray}
\end{subequations}
Here, a prime denotes differentiation with respect to $y$. 
Neglecting the diffusive terms on the right hand side and eliminating $\hat{\rho}$, we obtain the Taylor-Goldstein equation
\begin{eqnarray}
\hat{\psi}''-k^2\hat{\psi}-\frac{U''}{U-c}\hat{\psi}+\frac{J}{(U-c)^2}\hat{\psi}=0.\label{TGeq}
\end{eqnarray}
If the stratification is non-uniform, the constant $J$ must be replaced by a function $JN^2$, where $N$ is the Brunt-V\"ais\"al\"a frequency. The stratification profile studied in \citet{sutherland1994turbulence}, \citet{sutherland1994internal} corresponds to $N^2=\tanh^2(y/R)$, where $R$ is a constant.
Supporting such a non-uniform stratification requires an additional forcing term in (\ref{tempdim}).

Throughout the paper, by linear stability analysis, we mean solving (\ref{visstab}) for the complex eigenvalue $c(k,J,H,Re,Pr)$ subject to the boundary conditions
$\hat{\psi}=\hat{\psi}'=\hat{\rho}=0$ at $y=\pm H$. The computational results were cross-checked using two independently developed codes based on the Chebyshev collocation method. 
The domain of the Chebyshev polynomials $\eta\in [-1,1]$ is mapped onto the physical domain $y\in [-H,H]$ through
\begin{eqnarray}
y=2\eta HB/[H (1-\eta^2)+2B \eta^2].\label{map}
\end{eqnarray}
When the parameter $B$ is smaller than $H/2$, the mapping clusters the collocation points near $y=0$. The algebraic eigenvalue problem obtained after discretisation is solved using either the LAPACK routine ZGGEV or the MATLAB eig function. 
We often write the eigenvalue as $c=c^R+ic^I$.
The growth rate is found by $\sigma=kc^I$; when it is zero, the perturbation is neutrally stable. The number of collocation points is chosen such that the eigenvalues are reliable to more than 6 decimal places.
%[comment on numerical resolution]
%\Kengo{Do we also need to add $\hat{\psi}'(\pm H) =0$ as well for no slip? [yes]}
%\Kengo{Is it $-k\Im(c)$ or $k\Im(c)$? [expand the exponent] I think if it takes it normal mode ansatz form as mentioned above $\tilde{\psi}=\hat{\psi}(y)e^{ik(x-c t)}$ then we probably need to be consistent. $ik[x-(Re(c) + iIm(c))t] = ikx -ikRe(c)t + kIm(c)t$. So $\sigma = k Im(c)$ should be the growth rate?}

Nonlinear travelling wave solutions (i.e. ECS) of (\ref{fulleq}) are sought using the Newton-Raphson code developed by \citet{deguchi2017scaling}, which employs the Chebyshev collocation method in $y$ and the Fourier Galerkin method in $x$. The phase speed of the travelling wave is also denoted by $c$; it is one of the unknowns to be determined as part of the solution. 
%A detailed description of the numerical algorithm can be found in Deguchi et al. (2013)
The same mapping function (\ref{map}) is employed. 
A typical computation retains 600 Chebyshev modes and 40 Fourier harmonics. For large-domain ECS, these numbers are increased to 1200 and 60, respectively. 
%[how many L and M used? Maximum?][Maximum is LL=1200, dimM=120 (for H=50), others is around LL=600, dimM=40-60.]

DNS are carried out using the well-tested spectral-element solver Semtex \citep{blackburn2019semtex}. In this work, the two-dimensional computational domain is partitioned into a collection of adjoining, edge-conforming quadrilateral elements. 
Our DNS typically use 800 spectral elements, with 8th order polynomial interpolation in each element. 
%[I suggest shorten; like this?] [yes that's good feel free to cut the redundant part]

The consistency between the Newton and DNS codes is thoroughly tested using the stable ECS discussed in sections 4 and 6.

%The governing equations (\ref{fulleq}) is solved numerically using Semtex, the spectral-element code designed by \citet{blackburn2019semtex}. In this framework, the two-dimensional computational domain is partitioned into a collection of adjoining, edge-conforming quadrilateral elements. Within each element, the solution is represented by nodal Lagrange polynomial interpolants, producing a piecewise-polynomial approximation that remains $C^{0}-$continuous across element interfaces \citet{Deville_Fischer_Mund_2002, karniadakis1991high, canuto2007spectral}. The code has been tested and verified in various problems, ranging from ....[Xueraos/Rungie simulation flows?] to unsteady jets in stratified flows (\citet{vu2026planar}). Our actual simulations are ran with higher resolutions, with up to 800 elements and 8 number of mesh points along the edge of each element.
%\CV{For Carl: Add details here}

\section{Baseline results}

Our main interest lies in the effects of $J$ and $H$ on the ECS. A standard approach for obtaining ECS is to perform bifurcation analysis from the neutral curve identified by linear stability analysis. However, no systematic linear stability study covering a broad parameter range of the stratified Bickley jet appears to be available, and we thus begin with such an analysis in section 3.1. 
In addition, bifurcation analysis of the unstratified Bickley jet has not been reported. Therefore we also present these results in section 3.2 as a baseline before discussing our main findings.

\subsection{Linear stability results for $H=10$}

\begin{figure}
    \centering
    \includegraphics[width=\linewidth]{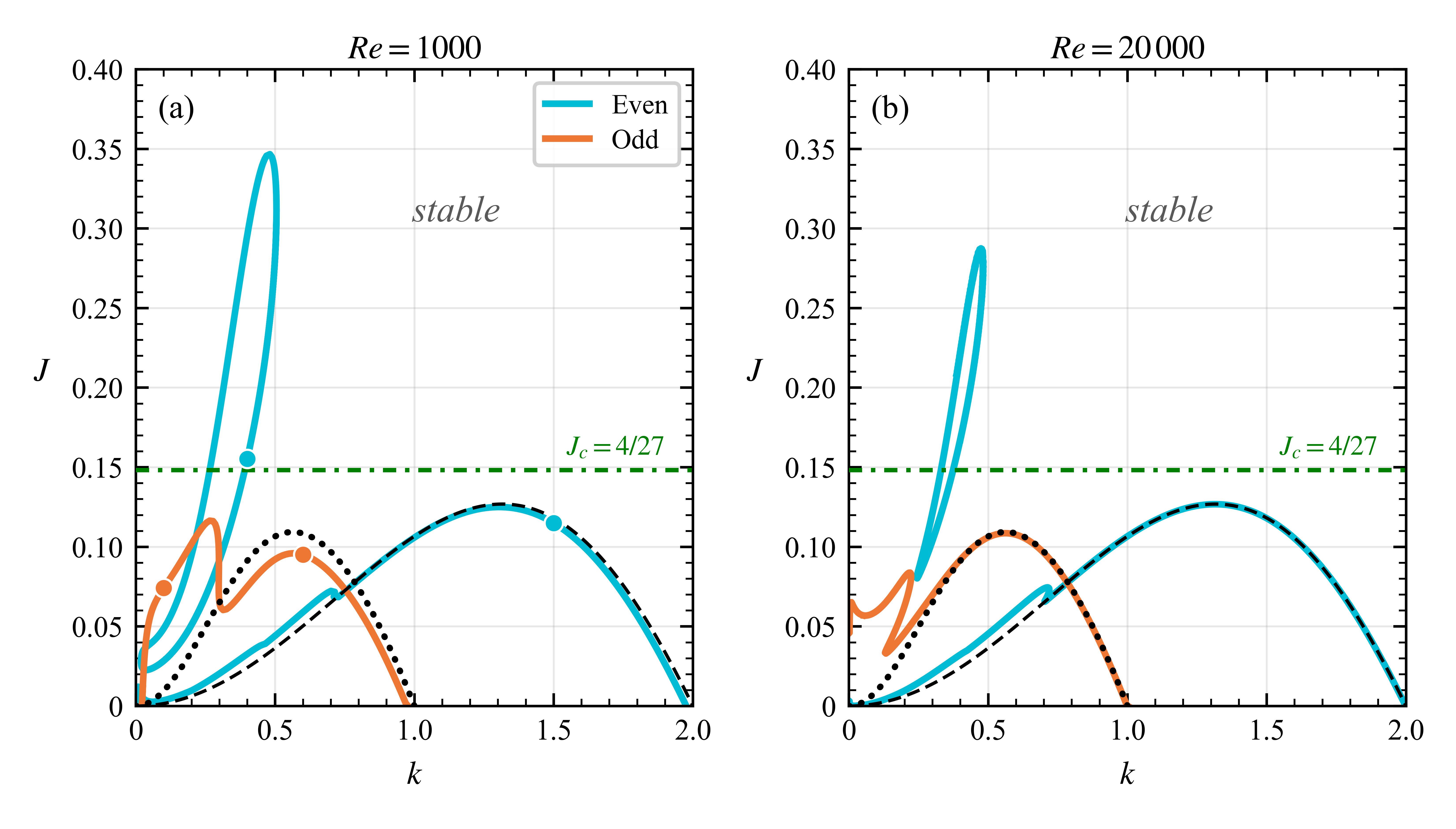}
    \caption{Neutral curves for $H=10$, $Pr=1$. The cyan and orange solid curves represent the even and odd modes, respectively, computed from (\ref{visstab}). The black curves are the inviscid results (\ref{invJ}); dashed/dotted correspond to the even/odd modes. The horizontal green dot-dashed lines indicate the Miles-Howard threshold $J=4/27$. (a) $Re=1000$, (b) $Re=20000$. The circles in panel (a) indicate the parameters used for figures 2 and 3.
    }
%    \Kengo{Do we need to shade the stable/unstable region here ? [no shading needed.]}
    \label{fig:linear stability boundary}
\end{figure}
This section presents linear stability analysis based on (\ref{visstab}) for $H=10$ and $Pr=1$.
Panels (a) and (b) of 
figure \ref{fig:linear stability boundary} show the neutral curves obtained at $Re=1000$ and $Re=20000$, respectively. The eigenfunctions are either even or odd functions, and the corresponding neutral curves are distinguished by cyan and orange lines. 
The neutral curves at the two values of $Re$ are qualitatively similar, except for small $k$, suggesting good asymptotic convergence even for $Re=1000$. 
We furthermore confirmed that the neutral curves for $Re=20000$ and $Re=30000$ are graphically indistinguishable for all $k$. 

Part of the asymptotic convergence can be explained by the analytic neutral solutions of the Taylor-Goldstein equation by \citet{drazin1966hydrodynamic}.
The black dashed and dotted curves in figure \ref{fig:linear stability boundary} are the neutral curves for
the even (sinuous) mode and the odd (varicose) mode,
\begin{eqnarray}
J=\frac{k^2(4-k^2)(9-k^2)}{225},~~~\text{and}~~~
J=\frac{k^2(1-k^2)(9-k^2)(3+k^2)^2}{9(3+5k^2)^2},\label{invJ}
\end{eqnarray}
respectively. 
Figure 2 compares the numerically obtained eigenfunction at $Re=1000$ with the analytic solution. 
Panels (a) and (b) show the even neutral mode obtained at $k=1.5$ (we denote $\hat{\psi}=\hat{\psi}^R+i\hat{\psi}^I$). 
The full numerical results computed with $(Re,H)=(1000,10)$ agree very well with the analytic solution
\begin{subequations}\label{HDeven}
\begin{eqnarray}\label{HDeven1}
\psi=
\left \{
\begin{array}{c}
(1-c-z^2)^{\mu}(1-z^2)^m \qquad \text{if}\qquad  |y|<y_c,\\
e^{-i\pi \mu}|1-c-z^2|^{\mu}(1-z^2)^m \qquad \text{if}\qquad  |y|>y_c,
\end{array}
\right .
\end{eqnarray}
where
\begin{eqnarray}
c=\frac{6+k^2}{15},\qquad \mu=\frac{3(4-k^2)}{2(6+k^2)},\qquad m=\frac{5k^2}{2(6+k^2)}.
\end{eqnarray}
\end{subequations}
The mapping
\begin{eqnarray}
z=\tanh y\label{maptanh}
\end{eqnarray}
transforms $y\in (-\infty,\infty)$ to $z \in (-1,1)$. Under this mapping, 
the base flow becomes $U=1-z^2$. 
Clearly the function (\ref{HDeven1}) is singular at the critical levels, $y=\pm y_c$, where $y_c>0$ is the location where $U(y_c)=c$ occurs.

Physically, the inviscid mode can be interpreted as a Kelvin-Helmholtz instability. 
As is well-known, the inviscid approximation breaks down within the critical layer of thickness $O(Re^{-1/3})$ centered at $y=y_c$.
A detailed analysis of the critical layer shows that a phase shift must be introduced in the outer wave across the critical level \citep{miles1961stability}.
%The phase shift has already been built into (\ref{HDeven1}), although its presence is not explicitly mentioned in \citet{drazin1966hydrodynamic}.
Due to the phase shift, even if the eigenfunction is normalised to be purely real in $|y|<y_c$, an imaginary part is produced for $|y|>y_c$ (here and hereafter the eigenfunctions are normalised so that $\hat{\psi}=1$ at which the location where $|\hat{\psi}|$ attains its maximum). 
\begin{figure}
    \centering
    \includegraphics[width=\linewidth]{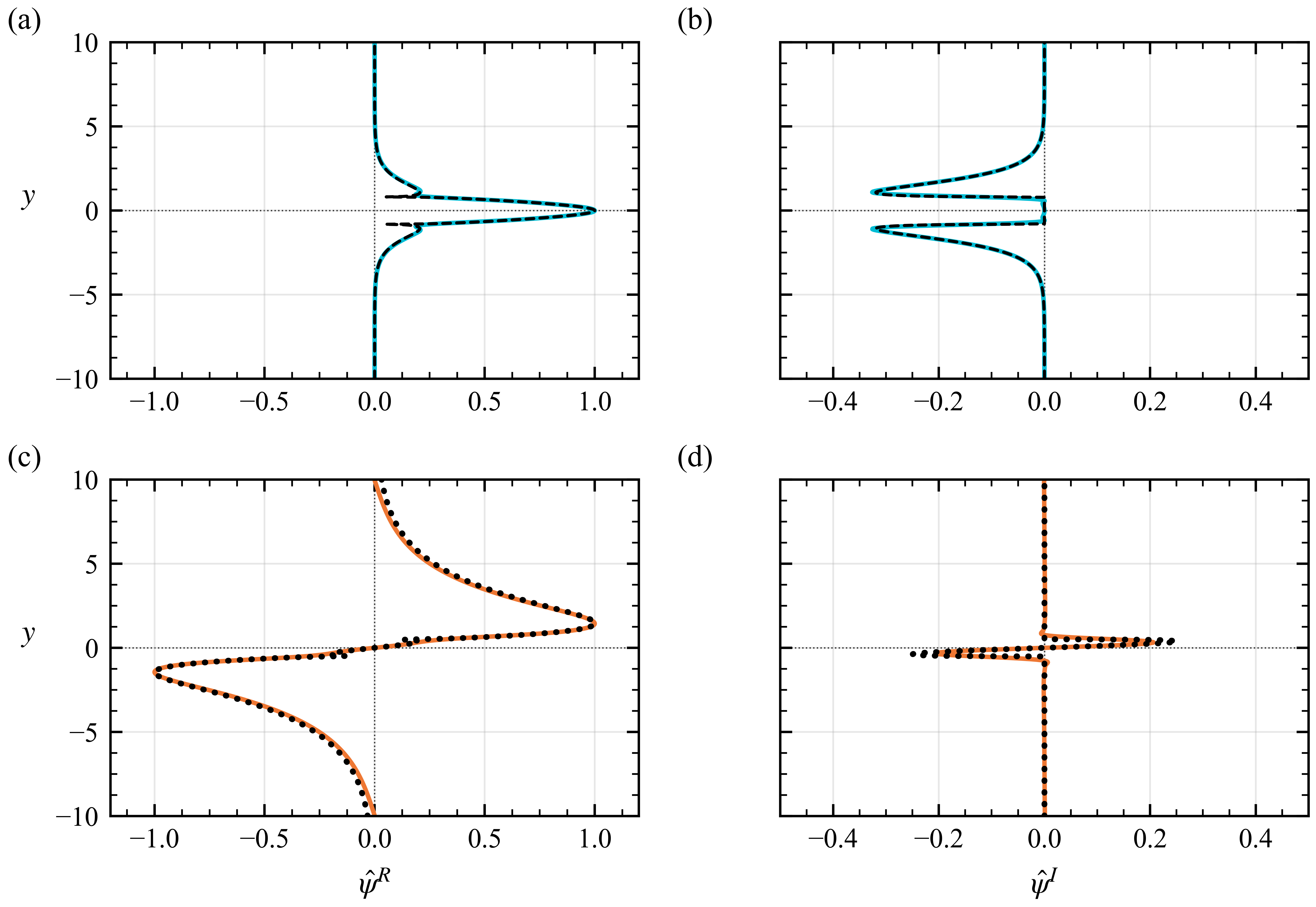}
    \caption{ Neutral eigenfunctions associated with the singular inviscid mode. 
The solid lines in panels (a) and (b) show the even mode numerically obtained at the rightmost circle in figure 1a ($k=1.5, J=0.114965$).  
The critical levels are at $y_c=\pm 0.81365$ ($c=0.5489567$).
The black dashed curves are the analytic solution (\ref{HDeven}) for the same $k$. 
%The critical levels are at $y_c=\pm ???$ (corresponding to $c=???$). 
Panels (c) and (d) show similar results for the odd mode at $k=0.6$. Solid line is used for the neutral point at $J=0.095096$ in figure 1a ($y_c=\pm 0.52458, c=0.768423$). 
The black dotted curves are the analytic solution (\ref{HDodd}).  }
    \label{fig:eigen_inviscid}
\end{figure}

A similar comparison for the odd neutral mode at $k=0.6$ is shown in panels (c) and (d). The corresponding analytic solution is 
\begin{subequations}\label{HDodd}
\begin{eqnarray}
\psi=
\left \{
\begin{array}{c}
(1-c-z^2)^{\mu}(1-z^2)^mz \qquad \text{if}\qquad  |y|<y_c,\\
e^{-i\pi \mu}|1-c-z^2|^{\mu}(1-z^2)^mz \qquad \text{if}\qquad  |y|>y_c,
\end{array}
\right .
\end{eqnarray}
where
\begin{eqnarray}
c=\frac{(3+k^2)^2}{3(3+5k^2)},\qquad \mu=\frac{3(1-k^2)}{2(3+k^2)},\qquad m=\frac{4k^2}{2(3+k^2)}.
\end{eqnarray}
\end{subequations}
All in all, the numerical results in figure 2 suggest that choosing $H=10$ and $Re=1000$ is sufficient to capture the inviscid singular modes in the unbounded domain.

The deviation from the results of \citet{drazin1966hydrodynamic} seen in figure 1b is due to the emergence of the viscous modes.
These modes are analogues to those identified by \citet{parker2020viscous} in a mixing layer with slip boundary conditions imposed at the walls. 
Consider the asymptotic expansions
\begin{subequations}\label{expRe}
\begin{eqnarray}
c=c_0+\epsilon c_1+\epsilon^2 c_2+\cdots,\\
\hat{\psi}=\psi_0(y)+\epsilon \psi_1(y)+\epsilon^2 \psi_2(y)+\cdots,\\
\hat{\rho}=\rho_0(y)+\epsilon \rho_1(y)+\epsilon^2 \rho_2(y)+\cdots,
\end{eqnarray}
\end{subequations}
assuming $\epsilon\ll 1$ and $c_0$ is real.
Substituting (\ref{expRe}) into (\ref{visstab}), the leading order solution simply satisfies the neutral Taylor-Goldstein equation with Dirichlet boundary conditions,
\begin{subequations}
\begin{eqnarray}
\mathcal{L}_0\psi_0=
\left (\mathcal{U}^2 \left (\frac{\psi_0}{\mathcal{U}}\right )' \right)'+(J-k^2\mathcal{U}^2)\frac{\psi_0}{\mathcal{U}}=0 \label{q0eq},\\
\psi_0=0\qquad \text{at} \qquad y=\pm H,
\end{eqnarray}
\end{subequations}
where $\mathcal{U}=U-c_0$. Here we write (\ref{q0eq}) in Sturm-Liouville form for later purposes. 
As shown in figure 3, the eigenfunctions of the finite $Re$ problem are well approximated by the neutral solutions of the Taylor-Goldstein equation. Panels (a) and (b) show the even neutral mode at $k=0.4$ and the odd neutral mode at $k=0.1$, respectively.
The phase speeds of these modes  
($c=1.133387$ in panel (a), $c=1.214997$  in panel (b)) are used to estimate the $c_0$ supplied to the Taylor-Goldstein equation. Since both values are greater than 1, there is no critical-layer singularity, and a standard Chebyshev collocation method can be used to find the eigenvalue $k^2$. Here,  
$c_0$ is tuned as $1.13394515$ in panel (a) and  $1.21447227$ in panel (b), so that the resulting eigenvalue gives $k\approx 0.4$ and $k\approx 0.1$, respectively. The associated eigenfunctions agree very well with those obtained without the asymptotic reduction. The former is purely real, but the latter involves a small imaginary part because of viscous effects at higher order. 
%\Kengo{swapped even and odd.}
The higher-order analysis determines the leading order contribution to the growth rate \textcolor{black}{via}
\begin{equation}
\sigma=\epsilon kc_1^I+\epsilon^2 kc_2^I+\cdots.
\end{equation}
\citet{parker2020viscous} applied slip boundary conditions on the walls and took $\epsilon=Re^{-1}$, so that the viscous terms in (\ref{visstab}) enter at $O(\epsilon)$. However, we shall show that when the walls are no-slip, %and their interaction with the perturbation is not negligible, 
the suitable choice is $\epsilon=Re^{-1/2}$. 
This is due to the Stokes boundary layer that develops near the walls, and the asymptotic analysis becomes similar to that discussed by \citet{duck1994linear} and \citet{deguchi2025instability}. 
A detailed asymptotic theory for our setup is given in section 5 and Appendix A.

\begin{figure}
    \centering
    \includegraphics[width=\linewidth]{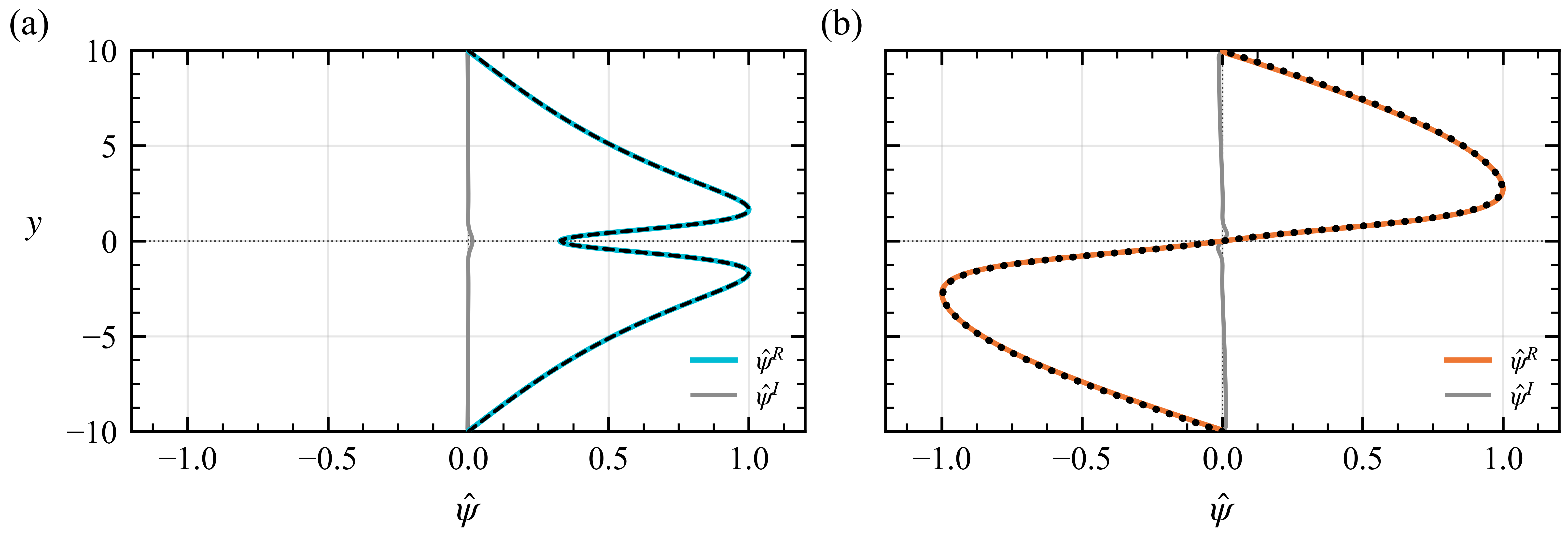}
    \caption{Neutral eigenfunctions associated with the viscous modes. The solid curves show the numerical solutions of (\ref{visstab}) at the two leftmost circles in figure 1a:  
(a) even mode at $(k,J)=(0.4,0.15535)$,
(b) odd mode at $(k,J)=(0.1,0.074207)$.
%The imaginary parts are very small and are therefore omitted. 
The black dotted/dashed curves are the neutral Taylor-Goldstein solution obtained at the same $k$.}
    \label{fig:eigen_viscous}
\end{figure}

%We also examined the behaviour of the growth rates in the vicinity of the neutral modes. 
%Unlike the inviscid modes, they depend on $Re$ and scale as $1/Re$. The asymptotic behaviour can be explained by computing $\sigma_1$, but we omit the details here. The solvability condition for equation (\ref{nextorder}) will be presented when it becomes necessary; see section 5.

%A more detailed asymptotic analysis of the viscous mode is presented in section 6.
%\citet{parker2019kelvin}

%[I think leading order inviscid neutral, plus next order viscous type analysis was performed earlier .. we need some survey]

%Analytic expression is unfortunately not possible .. but ...

As shown by \citet{hazel1972numerical}, the local Richardson number $Ri(y)=J/(U')^2$ attains a minimum value of $(27/16)J$. Thus, according to the Miles-Howard theorem, there is no inviscid instability above the critical value $J=4/27$ (indicated by the green dot-dashed lines in figure 1). The viscous even mode instability appears above this threshold. 
This behaviour again shares similarities with the viscous modes identified by \citet{parker2019kelvin}.

\subsection{Bifurcation analysis for $J=0$}
Seeding the Newton iteration with the neutral eigenfunction at a suitable amplitude results in convergence to a nontrivial solution. Once a solution has been obtained, numerical continuation using the Newton method can be employed to trace out the bifurcation diagram. Note that, throughout the paper, all ECS that converge under the Newton method are travelling waves.

\begin{figure}
    \centering
    \includegraphics[width=\linewidth]{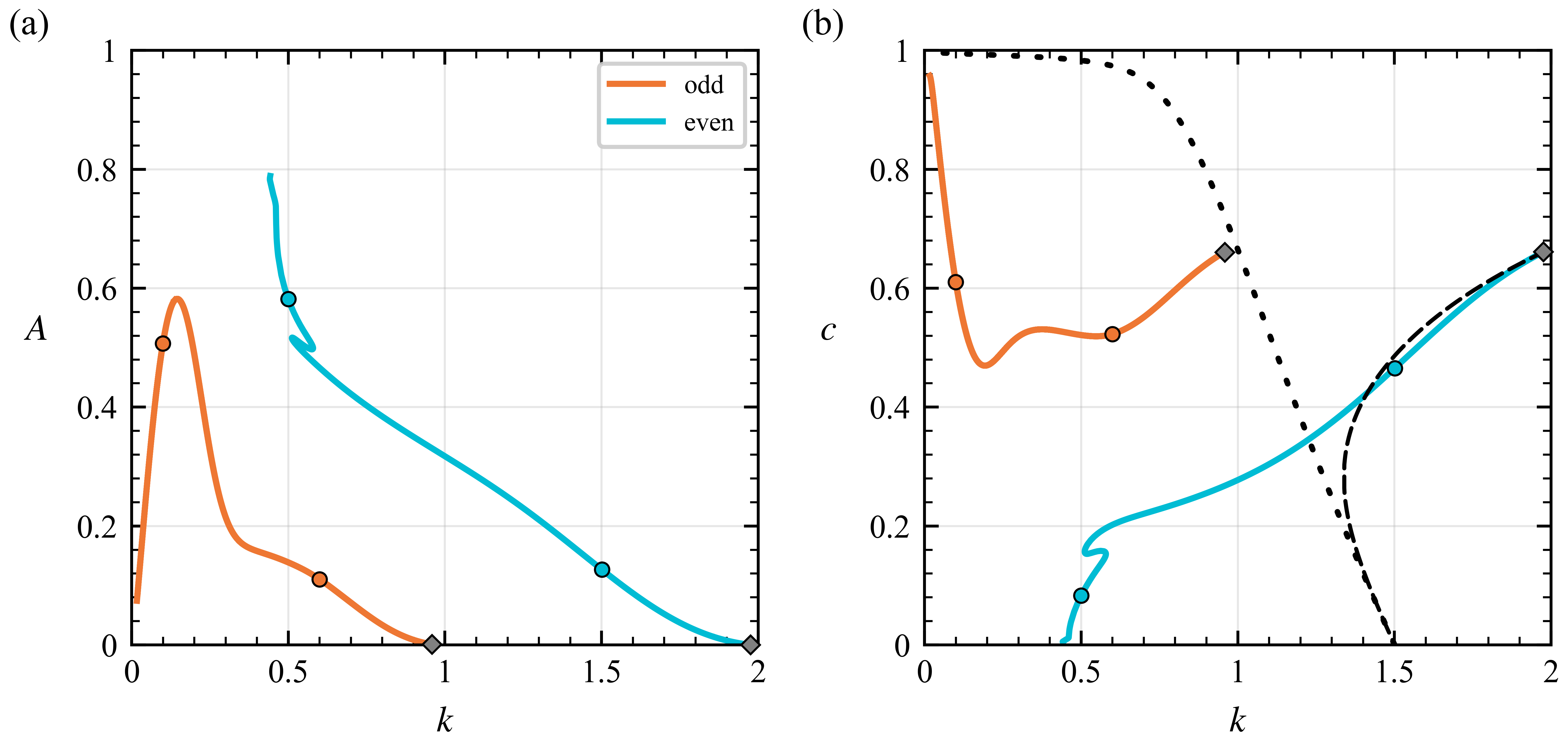}
    \caption{
    Bifurcation diagram obtained for the unstratified Bickley jet ($J=0$). The thick cyan and orange curves are the even and odd solution branches obtained with $(H,Re,Pr)=(10,1000,1)$. 
    %The black dashed curves show the same solution branches computed in the extended domain ($H=50$). 
    Bifurcation points are indicated by the grey diamonds. 
    (a) Amplitude $A$ defined in (\ref{amp}); (b) phase speed $c$. The black curves are the predictions of the nonlinear critical layer theory by \citet{benney1969new}; the dashed and dotted curves correspond to the even and odd modes, respectively. 
    %[indicate the bifurcation points by grey diamond][not plotted solution, bifurcation point (linear neutral)][manipulate data so that dashed look more cleanly]
    %[half of the diamonds is hiding?]
    %[Benney: black. dashed for even, dotted for odd][rerunning, the jump is not real]
    %[remove H=50, Re=20000][even odd 1000, same cyan orange format as before. ][indicate the solutions you created the flow field (fig 5, fig 6)][add Benney in (b)]%The green dot-dashed curves in panel (b) 
    %nonlinear results at $J=0$. 
    }
    \label{fig:nonlinear results J=0}
\end{figure}

Figure 4 shows the bifurcation diagram for $J=0$. Panel (a) uses the amplitude 
\begin{eqnarray}
A=\left (\frac{1}{2}\int^H_{-H}(\Delta\overline{u})^2dy \right)^{1/2},\label{amp}
\end{eqnarray}
based on the mean flow distortion $\Delta\overline{u}=\overline{u}-U$. Here $U$ is the base flow defined in (\ref{base}) and $\overline{u}$ is the total mean flow
\begin{eqnarray}
\overline{u}(y)=\frac{k}{2\pi}\int^{2\pi/k}_0u\, dx.
\end{eqnarray}
%[notation changed]
The odd and even mode solution branches bifurcate from the base flow at around $k=1$ and 2, respectively (grey diamonds). 
Panel (b) shows the corresponding phase speed plot. 
The value of $c$ at the linear neutral point is close to $c=2/3$, which is determined by the inflection point of the base flow \citep{savic1941acoustically, savic1943symmetrical, tatsumi1958stability}. %\textcolor{orange}{}\textcolor{black}{(Savic 1941, Tatsumi \& Kakutani 1958)}. 

\begin{figure}
    \centering
    \includegraphics[width=\linewidth]{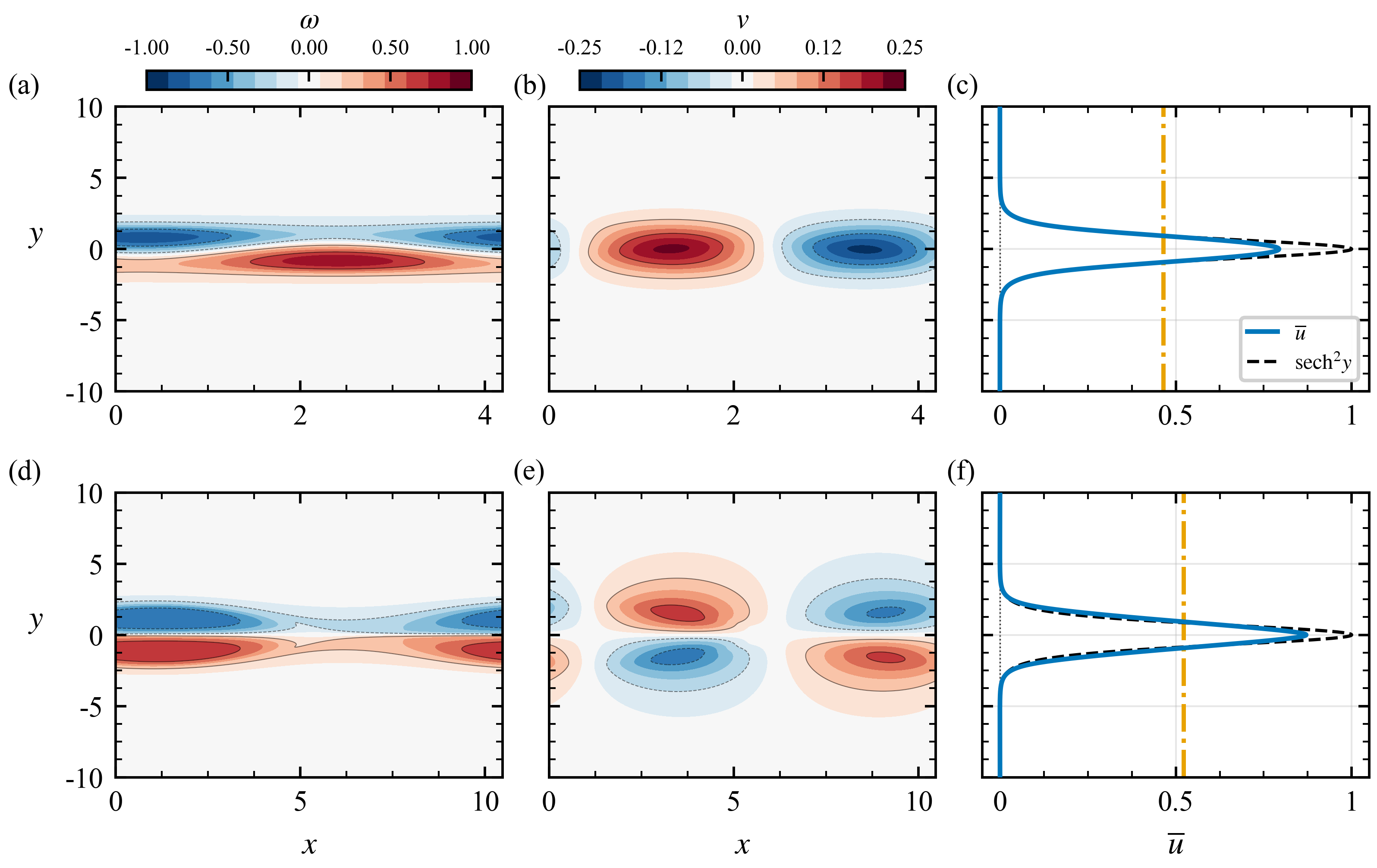}
    \caption{The flow field of the ECS at near the bifurcation point in figure 4. Panels (a,d), (b,e) and (c,f) show $\omega$, $v$, and $\overline{u}$, respectively. %The circles in (c,f) indicate the critical levels. [replace circles to gold yellow line]
    In panels (c,f), the yellow dot-dashed line indicates the phase speed $c$. %[remove the legend for the critical level]
    (a-c) Even mode at $k=1.5$; (d-f) odd mode at $k=0.6$. %[critical levels, also change $\overline{U}$ to $\overline{u}$]
    %
   % Vorticity k = 1.5 even and 0.6 odd , J=0. 
    %[ you cannot share the x axis? or not plotting one period]
    %[why panel b looks asymmetric?][I think you are including 0 contour. change colourbar definition]
    %[I think it would be good to use discrete colourmap with odd numbers of colours but we can tune later][also need to set max of the colourbar][check the parameters .. I requested even 1.5, odd 0.6 here, even 0.4 (0.5 is ok), odd 0.1 next fig][hat not needed]
    }
    \label{fig:k1p5_k0p5even_k0p4_odd}
\end{figure}

\begin{figure}
    \centering
    \includegraphics[width=\linewidth]{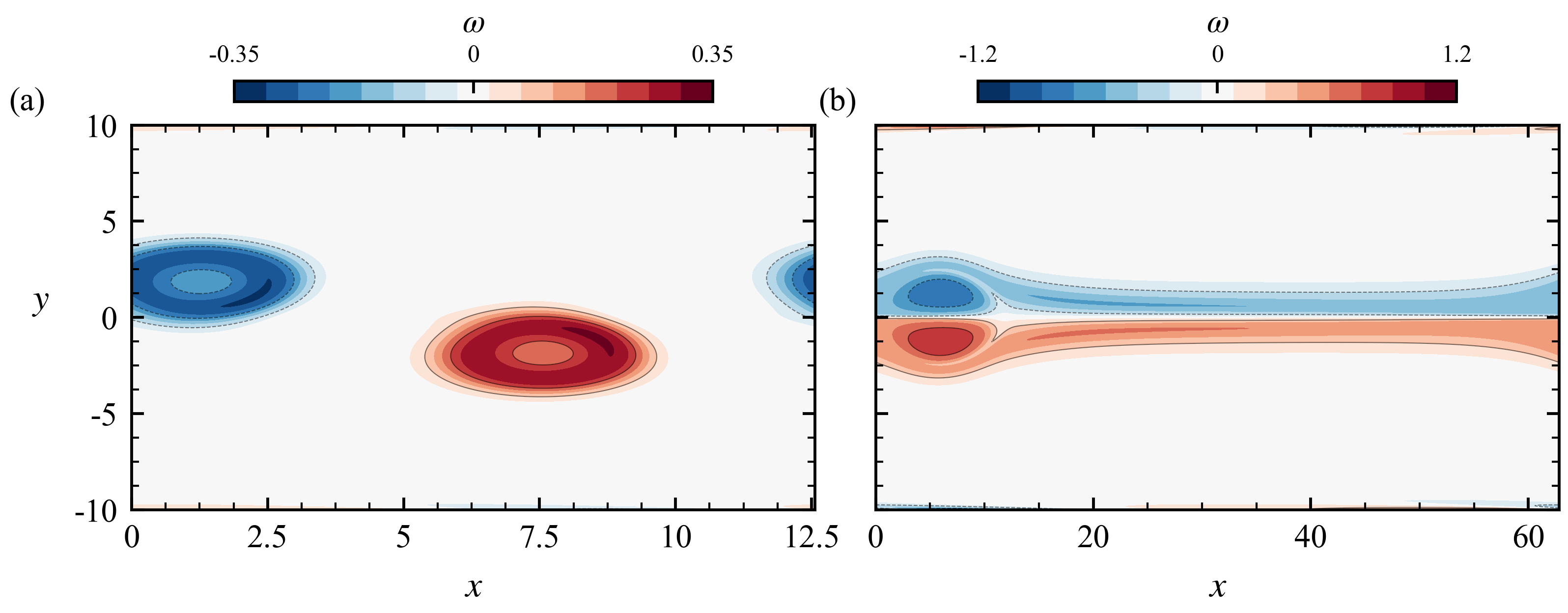}
    \caption{Vorticity of the relatively long wavelength ECS obtained in figure 4. 
    (a) Even mode at $k=0.5$; (b) odd mode at $k=0.1$.
    %Vorticity k = 0.1 odd and 0.5 even , J=0.
    }
    \label{fig:k1p5_k0p5even_k0p4_odd}
\end{figure}

Figure 5 depicts the flow field of the nonlinear solutions not far from the bifurcation points. 
There are three typical formats used for the flow field plots presented here and hereafter. 
%throughout the remainder of the paper.
Panels (a) and (d) show the vorticity $\omega=\partial_yu-\partial_xv$, which usually best illustrates the overall flow structure. Indeed, those panels clearly demonstrate why the even and odd modes are often referred to as the sinuous and varicose modes, respectively. Panels (b) and (e) show the normal velocity $v$. This type of plot is well suited for visualising the fluctuation field, as the $x$-averaged normal velocity is zero identically. 
Panels (c) and (f) show the streamwise mean flow $\overline{u}$. The critical levels can be found by locating the points at which $\overline{u}=c$. %[show this in the figure]

Figure 4b includes the prediction of $c$ from the nonlinear critical layer theory of \citet{benney1969new}. The theory assumes that nonlinear effects  dominate the critical layer structure. In contrast to the viscous critical layer, the critical layer has thickness %$O(Re^{-1/6})$
\textcolor{black}{depending on the wave amplitude,} and no phase jump occurs in the \textcolor{black}{leading order} outer solution.
We impose the zero phase jump condition on the Taylor-Goldstein equation (\ref{TGeq}) to compute the theoretical phase speed. 
Unfortunately, the agreement between the theoretical predictions and the ECS is mediocre for the even mode and poor for the odd mode. This might be not surprising, as the Reynolds number considered here is not very large (for example, \citealt{deguchi2018bifurcation} used Reynolds number $10^6$ to check the asymptotic convergence of ECS in plane Poiseuille flow). 
The theory by \citet{benney1969new} assumes that cat's eye vortices form within the critical layer. 
The flow field in figure 5a indeed exhibit such a structure around the critical level. 
That said, the vortices are not confined within the critical layer, and the two vortex layers appear to interact directly with each other.

We note that 
the nonlinear critical-layer theory
was extended to stratified shear flows by \citet{kelly1970nonlinear, maslowe1972generation, maslowe1973finite}. However, in view of the comparison in figure 4b, we conclude that these theoretical results are not relevant to most of the parameter range explored in this paper.

Figure 6 shows the solutions at longer wavelength (i.e. smaller $k$). The nonlinear effects associated with the vortical structures become stronger, leading to the formation of localised, large amplitude vortices around the \textcolor{black}{jet} centreline. 
The perturbations contain Fourier components that vary slowly in the streamwise direction and decay slowly towards the walls. Consequently, the interaction between the coherent structures and the walls cannot be neglected.
%For the even mode, localised regions of nearly constant vorticity emerge. 
%For the odd mode, a dipolar vortex structure emerges. 
%; similar vortical structures have been observed in various settings
%, for example in intrusion problems.
%; these show some resemblance to Jupiter's Great Red Spot. 

At several points on the bifurcation diagram, we increased $Re$ and $H$ to 20000 and 50, respectively. 
For $k\gtrsim 0.6$, the solution curves remain largely unchanged, while for smaller $k$, they become increasingly sensitive to $H$. 
This observation motivates us to investigate the large $k$ and small $k$ regimes separately. Section 4 focuses on $k=1.5$, where the solution behaviour is relatively simple. With the corresponding periodic box size, DNS can be performed efficiently while maintaining reliable resolution. 
Section 5 examines the long-wave regime with $k=0.1$. A DNS based parameter search is prohibitively difficult in this regime, and thus the Newton method is employed. 
In addition, we study the case $k=0.4$ in section 6. At this wavenumber, the viscous instability survives even under strong stratification (see figure 1), and the corresponding ECS is therefore of particular interest.

%The theory assumes that, except within the critical layer nonlinearity is negligible. Although this assumption is justified by a large Reynolds number asymptotic analysis, it is not strictly satisfied by our finite $Re$ solutions.

%The base flow has $A=0$. 
%The cyan and orange curves are the even and odd solution branches for $H=10, Re=1000$.

%We need to discuss if we include Re=20000

%Figure: Flow field plots at $k=1.5,0.5$ for even.$k=0.4, 0.1$ for odd.

% \begin{figure}
%     \centering
%     \includegraphics[width=\linewidth]{figures/velocity_vorticity_k1p5_J0_even.png}
%     \caption{Vorticity k = 1.5, J=0, even. 
%     }
%     \label{fig:k1p5_j0}
% \end{figure}

% \begin{figure}
%     \centering
%     \includegraphics[width=\linewidth]{figures/velocity_vorticity_k0p5_J0_even.png}
%     \caption{Vorticity k = 0.5, J=0, even. 
%     }
%     \label{fig:k0p5_j0}
% \end{figure}

% \begin{figure}
%     \centering
%     \includegraphics[width=\linewidth]{figures/velocity_vorticity_k0p4_J0_odd.png}
%     \caption{Vorticity k = 0.4, J=0, odd. 
%     }
%     \label{fig:k0p4_j0}
% \end{figure}

% \begin{figure}
%     \centering
%     \includegraphics[width=\linewidth]{figures/velocity_vorticity_k0p1_J0_odd.png}
%     \caption{Vorticity k = 0.1, J=0, odd. 
%     }
%     \label{fig:k0p1_j0}
% \end{figure}

%\begin{figure}
%    \centering
%    \includegraphics[width=\linewidth]{figures/velocity_vorticity_k0p1_J0_odd_h50.png}
%    \caption{Vorticity k = 0.1, J=0, odd, $H=50$. [this is not needed, will comment out]
%    }
%    \label{fig:k0p1_j0}
%\end{figure}

%For small k try $H=10,50$.

\section{Short wavelength regime}

\begin{figure}
    \centering
    \includegraphics[width=\linewidth]{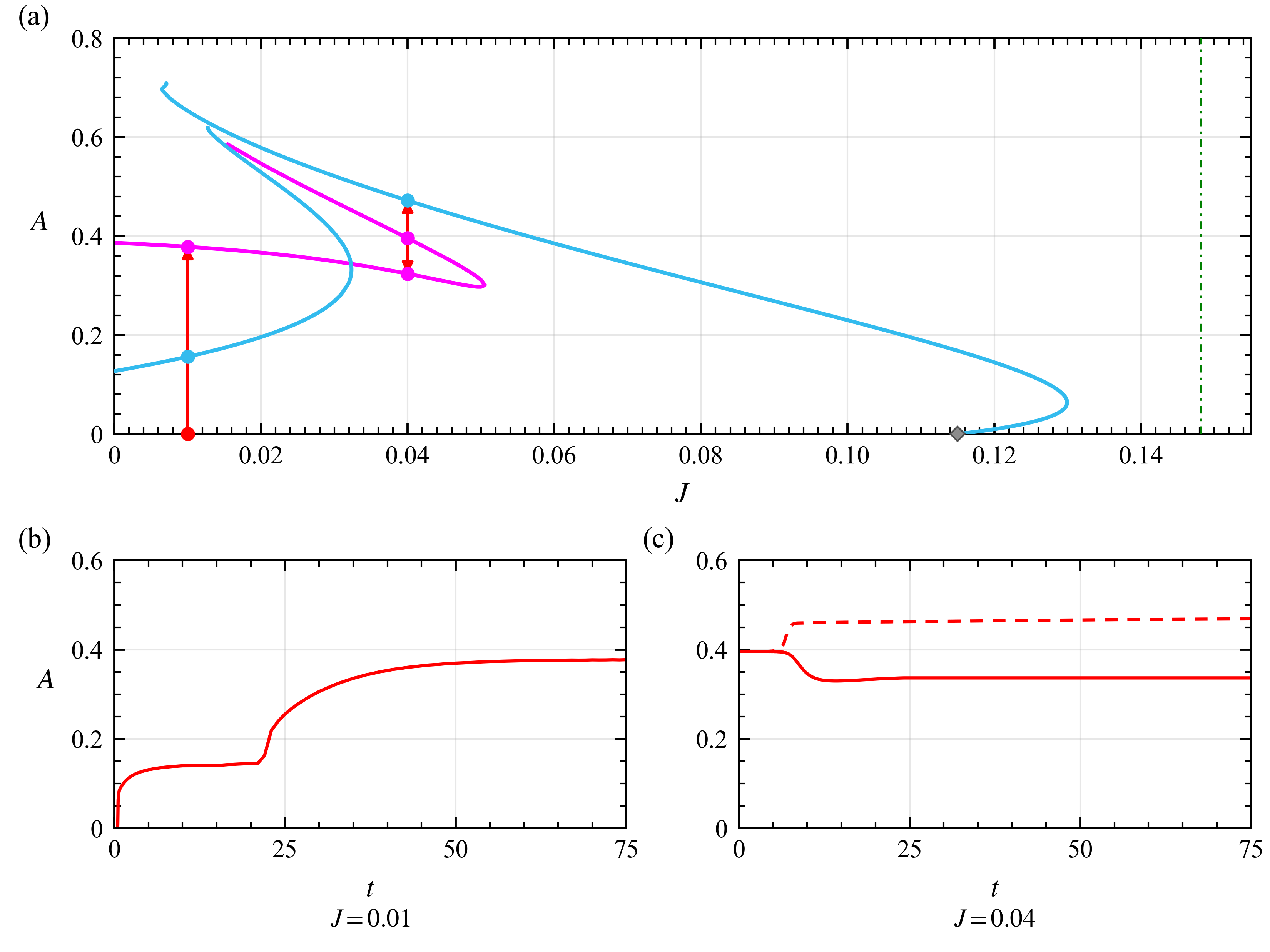}
    \caption{
    Nonlinear computations performed at $k=1.5$, with $(Re,Pr)=(1000,1)$. %, as in figure 4. 
    (a) Bifurcation diagram of travelling wave solutions obtained by the Newton method. The cyan and magenta curves are the even mode and asymmetric solution branches, respectively. The dot-dashed vertical green line indicates $J=4/27$. Arrows indicate the transition roots observed in the DNS in panels (b) and (c).
    (b) DNS result at $J=0.01$, starting from an $O(10^{-8})$ 
    perturbation of the base flow. 
    %[remove blue dashed] 
    %The red solid and cyan dashed curves represent the results obtained by
    %and from the even mode solution, respectively. 
    (c) DNS results at $J=0.04$. 
    The initial conditions are the lower branch asymmetric solution with \textcolor{black}{small random perturbations.}
    %\textcolor{black}{[use red curves in panel c]} 
    %[circles are too small][updated, but do you want a bit bigger?][looks ok]
    }
    \label{fig:short wavelength results}
\end{figure}
%\Kengo{Caption: did we mean small random perturbations or a small random perturbation? [fixed]}
Figure 7a shows the bifurcation diagram obtained for $k=1.5$. The two cyan curves are the even mode branch. One of them is continued from the $J=0$ solution in figure 4, while the other is obtained by bifurcation analysis around the neutral curve. They belong to the same solution branch; although we are unable to connect them at $k=1.5$, the connection can be found by drawing bifurcation diagrams at larger values of $k$. \textcolor{black}{In figure 7a, the bifurcation from the linear critical point is subcritical. We tried to continue the solution branch beyond the Miles-Howard threshold by varying $k$, but were unsuccessful.}

To link the transition process and the ECS, we performed several DNS. 
We first set $J=0.01$ and initialise the DNS with a small random perturbation added to the base flow. 
The time evolution of the amplitude is plotted in figure 7b. 
The linearly unstable modes grow exponentially, as expected, after which $A$ exhibits a plateau around $t=20$.
This plateau corresponds to a temporal approach towards the even mode solution, marked by the cyan circle in figure 7a.
%
%Although $A$ exhibits a temporary plateau around $t=20$, 
A second transition soon follows. The final state possesses neither even nor odd symmetry; we therefore refer to it as the asymmetric mode. 
In figure 7a, the asymmetric mode solution branch is plotted in magenta. The final state of the DNS lies on its lower branch. 

Additional DNS runs confirmed that the lower asymmetric solution branch is stable over the range of $J$ shown in the bifurcation diagram, while its upper branch counterpart is unstable.
The even mode branch also maintains stability, but only over the range $J\gtrsim 0.04$ up to the saddle node bifurcation at $J=0.129$. 
At $J=0.01$, we confirmed that DNS initiated from this branch results in a transition to the lower branch asymmetric solution.

%For example, when the DNS is initialised by the even mode solution indicated by the cyan circle [plot this], the time integration eventually settles onto the asymmetric solution, as shown in figure 7b (cyan curve).
At $J=0.04$, there are two stable states; the lower branch asymmetric solution and the even mode solution. 
Figure 7c demonstrates that both states can be accessed by adding a small perturbation to the upper branch asymmetric solution. 
Therefore, the latter solution lies on the boundary between the basins of attraction of the two stable solutions. 
In the terminology of shear flow studies \citep{toh2001regeneration, skufca2006edge} it is an edge state.   
%; see the magenta curves in . 

%We also performed DNS initialised with a small random perturbation (the red curve). The final state is an even mode solution, unlike the $J=0.01$ case. Understanding the change in the transition process between $J=0.01$ and 0.04 would require a global bifurcation analysis, which is beyond the scope of the present study.

\begin{figure}
    \centering
    \includegraphics[width=\linewidth]{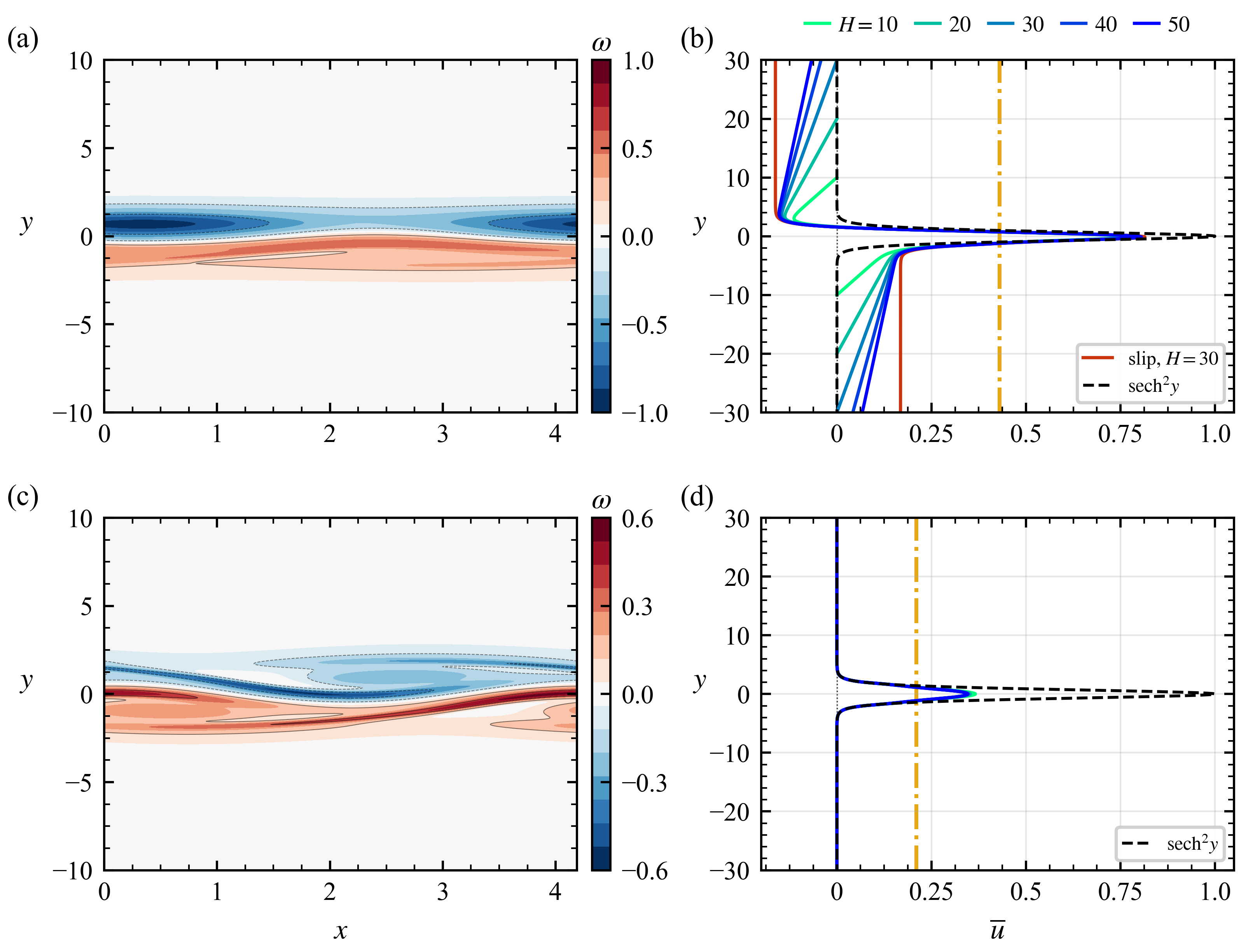}
    \caption{Flow fields of the stable ECS obtained at $J=0.04$. (a,b) asymmetric solutions; (c,d) even mode solutions.
    Panels (a) and (c) represent the vorticity $\omega$ of the stable solutions obtained at $H=10$. 
    Panels (b) and (d) show the mean flow $\overline{u}$ for $H=10,20,30,40,$ and 50. The red curve in panel (b) is obtained by imposing slip boundary conditions at $H=\pm 30$.  %[label not needed in a,c][use the same x range in b,d][colour map needs to be fixed to the same style as fig 5]
    %Mean $U$ varying $H$ for $k=1.5$, $J=0.04$ lower and asym vorticity. [use H=30 for slip]
    %[Update nicer values of H later, slip varying H running][mean density?]
    }
    \label{fig:varyingH_k1.5_j004}
\end{figure}
The flow fields of the stable solutions obtained at $J=0.04$ are shown in figure 8. 
Panels (a) and (c) compare the vorticity plots, with the asymmetry in the former clearly visible.
Interestingly, the asymmetric solution maintains a non-vanishing mean flow in the far field, as shown in figure 8b. This effect persists even if we increase $H$. The different lines in the figure were obtained by varying $H$ of the solution (see figure 9) and selecting representative results. 
The profiles above and below the jet are \textcolor{black}{almost} straight lines with the same slope. This follows from the momentum transport conservation
\begin{eqnarray}
Re^{-1}(\Delta\overline{u})'-\overline{\tilde{u}\tilde{v}}=M,\label{MMM}
\end{eqnarray}
where $\tilde{\mathbf{u}}=[u-\overline{u},v]$ is the fluctuation velocity components, and the overline denotes a streamwise average.
It can be shown from the $x$-component of (\ref{momentumeq}) that the momentum $M$ appearing on the right hand side must be a constant. 
Therefore, the slope $\overline{u}'$ above and below the jet coincide when the fluctuation components are negligible outside the jet (note that $\overline{u}'\approx (\Delta \overline{u})'$ there).
%The argument here also demonstrates that, for a symmetric mean flow, $\overline{u}'$ vanishes in the far field, as shown in figure 8d, provided that the coherent structures remain confined to the jet.
As $H$ is increased, the slope becomes less steep as the linear profile adjusts to satisfy the no-slip boundary conditions at the walls (i.e. the magnitude of $M$ decreases). 

Figure 8d \textcolor{black}{illustrates} the mean flow of the even mode in the same format as figure 8b. The mean flow profiles for $H=10,20,\dots,50$ almost collapse onto a single curve. 
There is no shear generated outside of the jet because 
$M=0$ when the mean flow is symmetric. \textcolor{black}{This conclusion follows} from the fact that the second term on the left hand side of (\ref{MMM}) vanishes at the walls, while the first term takes opposite signs at the upper and lower walls.

\begin{figure}
    \centering
    \includegraphics[width=\linewidth]{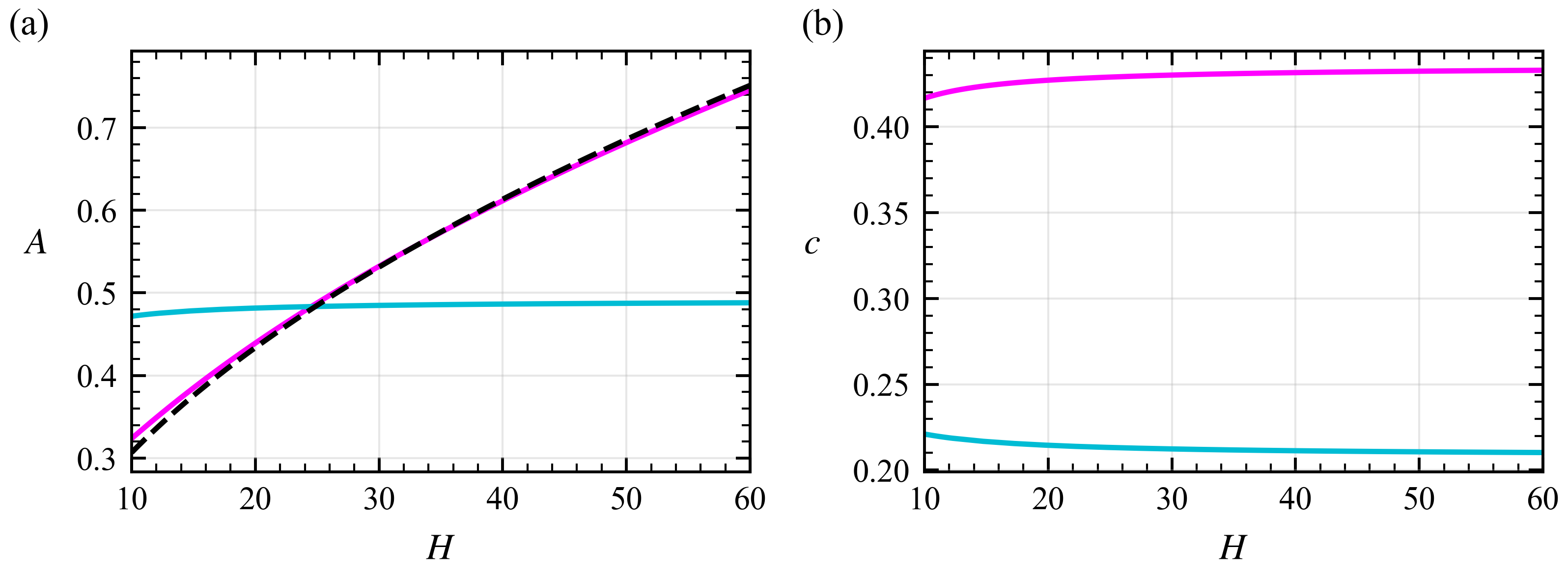}
    \caption{Bifurcation diagrams obtained by increasing $H$ from the stable solutions at $J=0.04$ shown in figure 7. The magenta and cyan curves represent the asymmetric and even mode solutions, respectively. The black dashed curve is the approximation (\ref{approx}). %[perhaps panel size is too large]
    %[remove legends]
    %Varying H for $k=1.5$, $j=0.04$ asym lower and sym stable. orange is the approximation curve. $V~0.168$[even cyan, asym magenta, approx black dashed][only upto 60]
    }
    \label{fig:placeholder}
\end{figure}
Figure 9b shows that $H\approx 30$ is sufficient to achieve convergence of the phase speed for both solutions. 
The coherent structure within the jet almost converges, but as we saw in the above, the mean flow part of the asymmetric mode does not. As a consequence, the value of $A$ for the asymmetric mode increases monotonically, as shown in figure 9a. 

The solution in the large $H$ limit can be better approximated by imposing slip boundary conditions.
Applying 
%\begin{eqnarray}
%\partial_y(u-U)=0,\qquad v=0
%\end{eqnarray}
\textcolor{black}{(\ref{slipBC}) at $y=\pm H=30$,} we confirmed that 
the corresponding ECS is stable and maintains asymmetric coherent structures within the jet, very similar to that seen in figure 8a. 
The ECS induces the mean flow indicated by the red curve in figure 8b, which behaves like a piecewise linear profile
\begin{eqnarray}
\overline{u}=
\left \{
\begin{array}{c}
((y/H)-1)\overline{u}_{\infty}\qquad \text{if}\qquad y>0,\\
((y/H)+1)\overline{u}_{\infty}\qquad \text{if}\qquad y<0.
\end{array}
\right .\label{plinearmean}
\end{eqnarray}
The constant $\overline{u}_{\infty}$ is estimated to be $0.168$ from the boundary values of the mean flow in figure 8b. %is obtained as half of the velocity difference of the mean flow between the upper and lower walls.
Plugging (\ref{plinearmean}) into (\ref{amp}), we obtain
\begin{eqnarray}
A=\overline{u}_{\infty}\sqrt{H/3}. \label{approx}
\end{eqnarray}
This is the dashed curve in figure 9a, which provides an excellent explanation of the behaviour observed in the numerical results.

\section{Long wavelength regime}

In section 3, we have already obtained an odd mode ECS at $k=0.1$ for the unstratified Bickley jet. A natural approach is therefore to increase $J$ from this solution, as will be done in section 5.1. Section 5.2 demonstrates that varying $H$ can induce mode switching between linear instability modes. Bifurcation analysis at different values of $H$ can thus lead to qualitatively different ECS, as we shall see in section 5.3.

\subsection{Momentum transport by internal gravity wave}
\begin{figure}
    \centering
    \includegraphics[width=\linewidth]{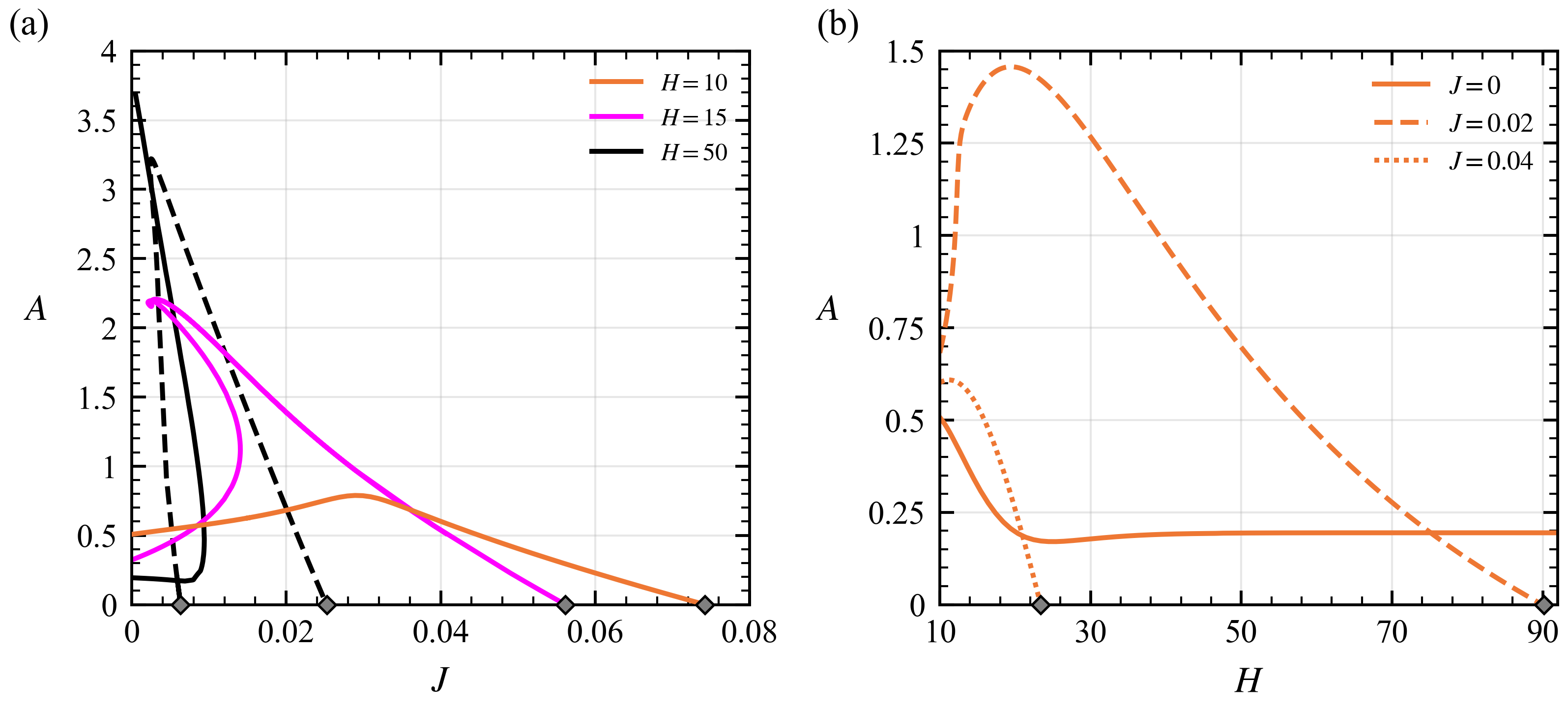}
    \caption{
    Bifurcation diagram of the odd-mode solution branches for $(k,Re,Pr)=(0.1,1000,1)$. (a) Results obtained by fixing $H$. For $H=50$, the two disconnected branches are denoted by different line styles. (b) Results obtained by fixing $J$. %[I think legend for dashed is not needed]%[panel b yrange (0,1.5)][grey diamond at the bifurcation points][also in panel a, there are 4 points]
    %[also extend beyond 90 slightly to include bifurcation point]
%    nonlinear results at $k=0.1$ for different branches.
    %[place black curves behind, magenta next, orange front][I edited this but seems not much change? ]
    %[J=0 branch varying H still running, but seems to converge very slowly.]
    }
    \label{fig:varyingH_k0p1}
\end{figure}
Our starting point is the odd mode solution at $(k,H,Re,Pr,J)=(0.1,10,1000,1,0)$ obtained in figure 4. 
Increasing $J$ from this solution while keeping $H=10$ produces the orange curve in figure 10a. The ECS returns to the base flow around $J=0.0742154$ without undergoing a turning point. 
However, increasing $H$ to 15 (magenta curve) introduces a fold in the solution branch. 
At even larger $H=50$, the solution branch splits into two parts, as shown by the black solid and dashed curves. 
%We are unable to continue the turned branch to the $J=0$ limit. 
Stratification appears to give rise to a solution with large $A$; no such solution exists when $J=0$.

Figure 10b is the bifurcation diagram obtained by increasing $H$ \textcolor{black}{with $J$ fixed.} The solid curve again starts from the odd mode solution obtained at $(k,H,Re,Pr,J)=(0.1,10,1000,1,0)$. 
In this unstratified computation, the solution curve converges to a constant for sufficiently large $H$. 
However, even a relatively small amount of stratification prevents this convergence, and the perturbation amplitude vanishes at finite $H$. 
For example, at $J=0.02$ and 0.04, the solutions return to the base flow at $H=90.166$ and 23.37426, respectively.

%This also reflects the fact that the flow spreads beyond the jet and interacts with the walls. 
%Also, figure 10a shows that as $H$ increases, the amplitude drops to zero at progressively smaller $J$.

\begin{figure}
    \centering
    \includegraphics[width=\linewidth]{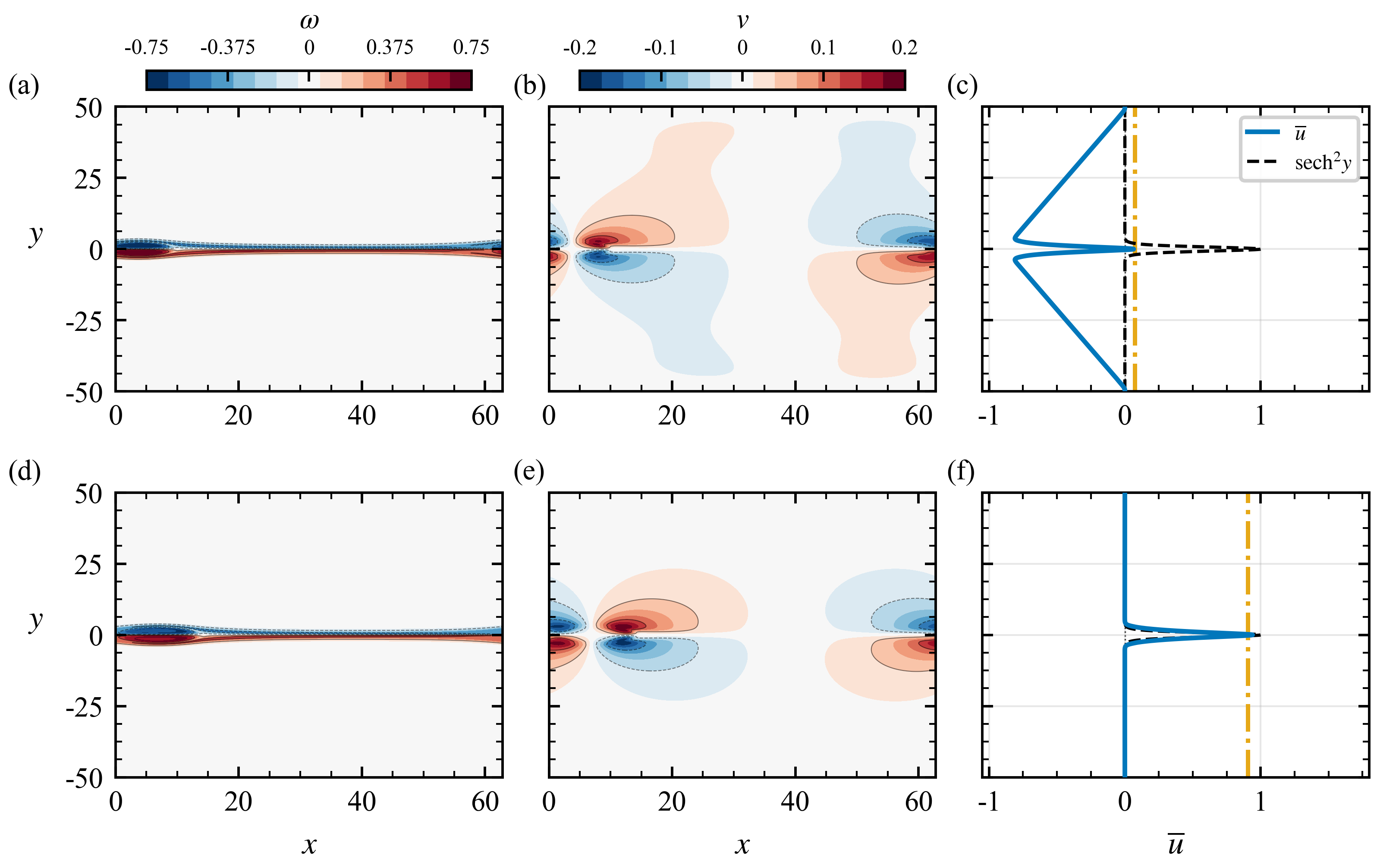}
    \caption{
    Flow field obtained from the black solid solution branch shown in figure 10a. The parameters are $(k,H,Re,Pr,J)=(0.1,50,1000,1,0.001)$. Same format as figure 5.
    }
    \label{fig:varyingH_k0p1}
\end{figure}

\begin{figure}
    \centering
    \includegraphics[width=\linewidth]{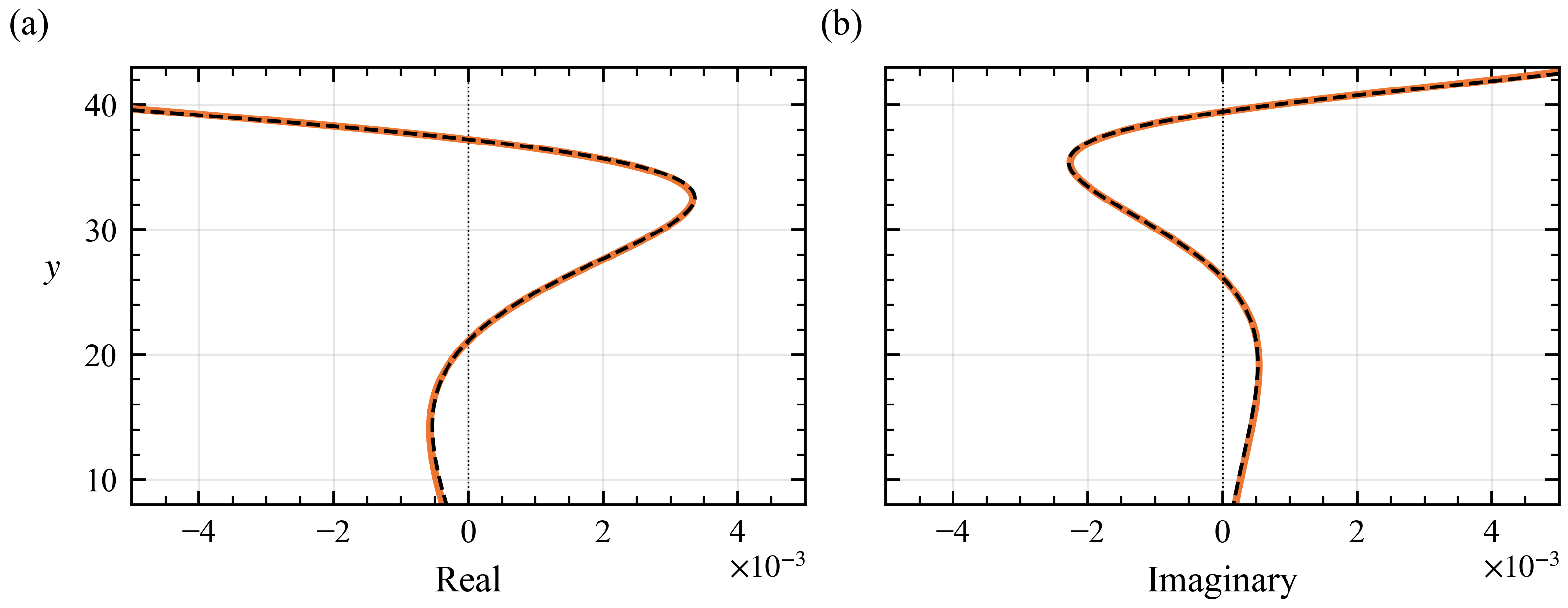}
    \caption{Confirmation that the outer jet flow structures observed in figure 11b are internal gravity waves. (a) Comparison between $(\hat{\psi}_1^R)''$ (orange solid curve) and $Q_1\hat{\psi}_1^R$ (black dashed curve). (b) The same comparison for $\hat{\psi}_1^I$. %[red to orange][remove x label, legends]
    %\Kengo{So we dont need $(\Psi_1'')^R$ ($ \times 10^{-3}$)? [just put Real/Imaginary][put rescaling $10^{-3}$ at the right bottom][looks ok]}
    %
%    Compare outer waves if it satisfies $\hat{\Psi}'' = Q\hat{\Psi}_1$. Figure 11d data [dashed is Qpsi?][do you have the similar plot for n=2? Note that $Q_n$ depends on n]
    }
    \label{fig:placeholder}
\end{figure}
The phenomena observed in those two panels are associated with the mean flow spreading beyond the jet. \textcolor{black}{The underlying physical mechanism responsible for mean-flow generation differs from that described in section 4 as we shall explain shortly.}
%, as now the mean flow is directly driven by the spreading fluctuations. 
Figure 11 shows the typical flow fields with and without such spreading.
The top and bottom panels correspond to the upper and lower branches of the black solid curve seen in figure 10a, respectively. 
The former large $A$ solution emits waves spanning the whole channel (panel (b)), which in turn generate a large scale mean flow that interacts with the walls (panel (c)). 
On the other hand, the structures within the jet are similar for both solutions, consisting of a localised dipole vortex structure.

%The small $A$ solution is localised within the jet, while the large $A$ solution develops large scale structures spanning the whole channel. 

The mean flow production mechanism can be understood by examining  (\ref{MMM}), as described below.
When $M=0$, the momentum transport by the mean flow is balanced by the Reynolds stress: 
\begin{eqnarray}
Re^{-1}(\Delta\overline{u})'=\overline{\tilde{u}\tilde{v}}=-\frac{k}{2\pi}\int^{\frac{2\pi}{k}}_0 (\partial_y\tilde{\psi})(\partial_x\tilde{\psi}) dx.\label{shearRS}
\end{eqnarray}
Here $\tilde{\psi}$ is the stream function for the fluctuations 
($\tilde{u}=\partial_y\tilde{\psi}, \tilde{v}=-\partial_x\tilde{\psi}$). 
Without loss of generality, $\tilde{\psi}$ admits the Fourier expansion
\begin{eqnarray}
\tilde{\psi}=\sum_{n=-\infty}^{\infty}\hat{\psi}_n(y)e^{ink(x-ct)}.
\end{eqnarray}
Splitting the coefficients into their real and imaginary parts as $\hat{\psi}_n=\hat{\psi}_n^R+i\hat{\psi}_n^I$, we obtain
\begin{eqnarray}
\overline{\tilde{u}\tilde{v}}=\sum_{n=1}^{\infty}2nk\{(\hat{\psi}_n^R)'\hat{\psi}_n^I-(\hat{\psi}_n^I)'\hat{\psi}_n^R\}.\label{RSt}
\end{eqnarray}

\begin{figure}
    \centering
    \includegraphics[width=\linewidth]{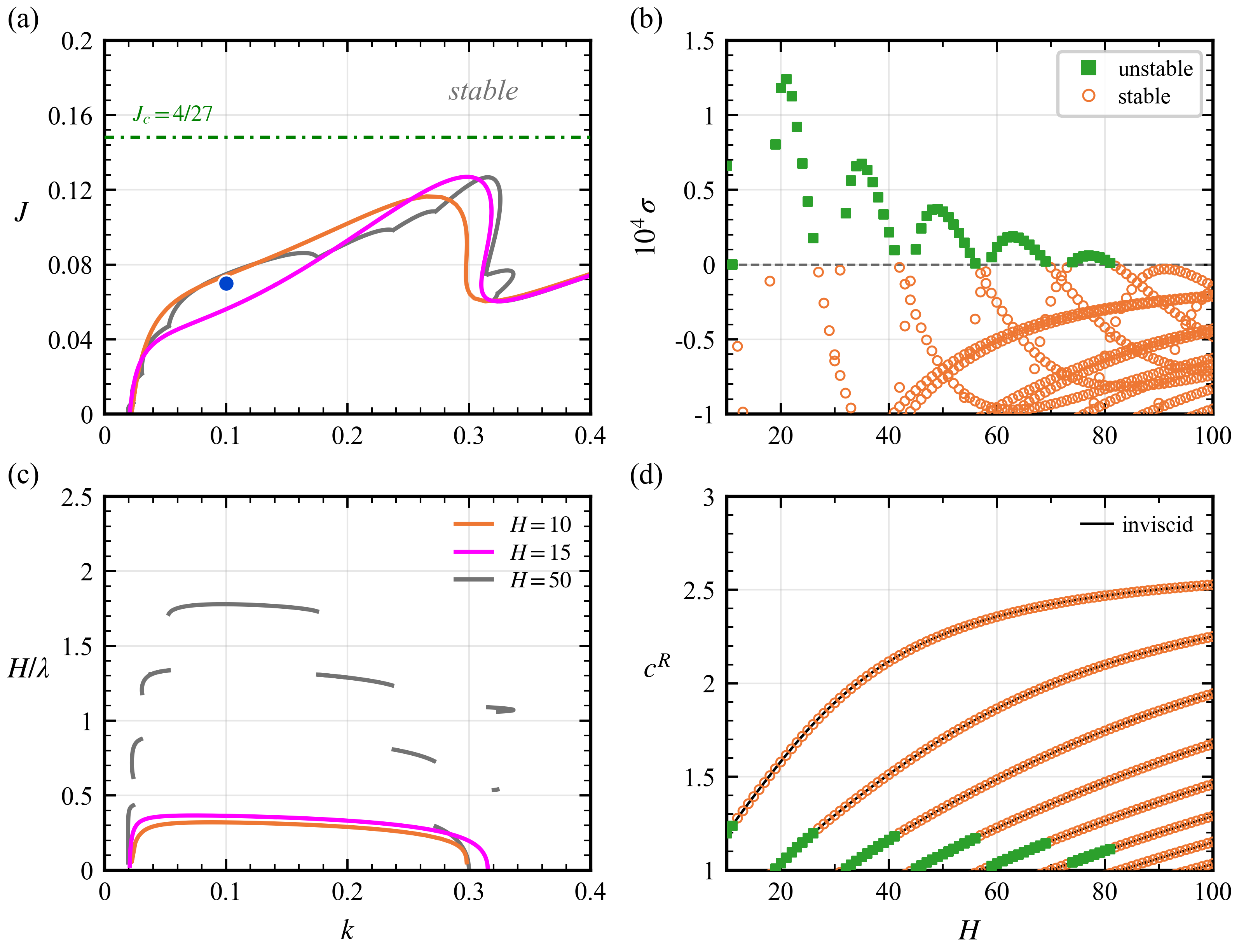}
    \caption{
Linear stability analysis of the long-wavelength regime. $(Re,Pr)=(1000,1).$ (a) Neutral curves in the $k$-$J$ plane. The circle indicates the parameter values examined in panel (b). (b) Dependence of the growth rates on $H$ at $(k,J)=(0.1,0.07)$. (c) The ratio of the domain half-height to the internal gravity wave wavelength along the neutral curve shown in panel (a). 
(d) The real part of the complex phase speed associated with the computation in panel (c). The black curves are the neutral eigenvalues of Taylor-Goldstein equation. %[panels b,d: place small tics at 30,50,..]%[H=50 purple][can we do H=15 a different color? maybe dark blue ?][Try H=50 grey][H=15 to magenta] %[top black curve truncated at H=90?][$c_r$ to $c^R$]
    %
    %fig 13: Linear neutral curves at odd mode $k=0.1$ \Kengo{Is this clear enough or should we use green square? [green square would be better. Remove triangles.]}
    }
    \label{fig:varyingH_neutral_curves}
\end{figure}
The fluctuations outside the jet observed in figure 11b are internal gravity waves. 
Evidence for this can be obtained by multiplying $\hat{\psi}_n^R$ by
\begin{eqnarray}
Q_n(y)=(nk)^2+\frac{\overline{u}''}{\overline{u}-c}-\frac{J}{(\overline{u}-c)^2}
\end{eqnarray}
and comparing the result with $(\hat{\psi}_n^R)''$. The good agreement obtained for $n=1$, as demonstrated in figure 12a, implies that $\hat{\psi}_1^R$ approximately satisfies the Taylor-Goldstein equation with the base flow replaced by the mean flow. 
Note that $Q_n$ has no singularity outside the jet. 
Likewise, $\hat{\psi}_1^I$ also approximately  satisfies the Taylor-Goldstein equation in the same sense; see figure 12b. 
Similar behaviour is observed for the higher harmonics, although the agreement is not as good as for $n=1$.
%, probably due to viscous effects. 

Now, there are several remarks to be made. First, for a single neutral Taylor-Goldstein solution without singularities, the right hand side of (\ref{RSt}) is zero. This is because the Taylor-Goldstein operator is real, and therefore the Fourier coefficient can be chosen to be purely real. However, vertical variation of the complex phase is possible when the waves are generated through a nonlinear interaction mechanism, such as that occurring within the dipole vortex. 
Second, if $(\hat{\psi}_n^R)''=Q_n\hat{\psi}_n^R$ and $(\hat{\psi}_n^I)''=Q_n\hat{\psi}_n^I$, the right hand side of (\ref{RSt}) is constant. This can be readily confirmed by Abel's identity for second order ordinary differential equations (the expression inside the curly brackets is the Wronskian of two solutions of the Taylor-Goldstein equation). By returning to (\ref{shearRS}), we can see why the mean flow profile in figure 11c becomes a linear function outside the jet.
Third, stable stratification plays a crucial role in this phenomenon. When $J=0$, the function $Q_n$ is dominated by the positive term $(nk)^2$, causing the waves to decay exponentially towards the walls \textcolor{black}{(as noted for linear stability problem by \citet{drazin1979normal}).}

As remarked in section 4, $M$ is indeed zero for a symmetric mean flow. The key to deducing this is that $(\Delta \overline{u})'$ vanishes at the wall. 
At first sight, the mean flow profile in figure 11c appears to contradict this, because of the linear profile induced by internal gravity waves. 
However, there is in fact a very thin near wall boundary layer (i.e. Stokes layer) within which $(\Delta \overline{u})'$ is reduced to zero at $y=\pm H$.

%Radiating internal gravity waves generated by the dipole structure can also be observed in DNS of the intrusion problem, as discussed in section 6.

\subsection{Radiating mode instability}
\begin{figure}
    \centering
    \includegraphics[width=\linewidth]{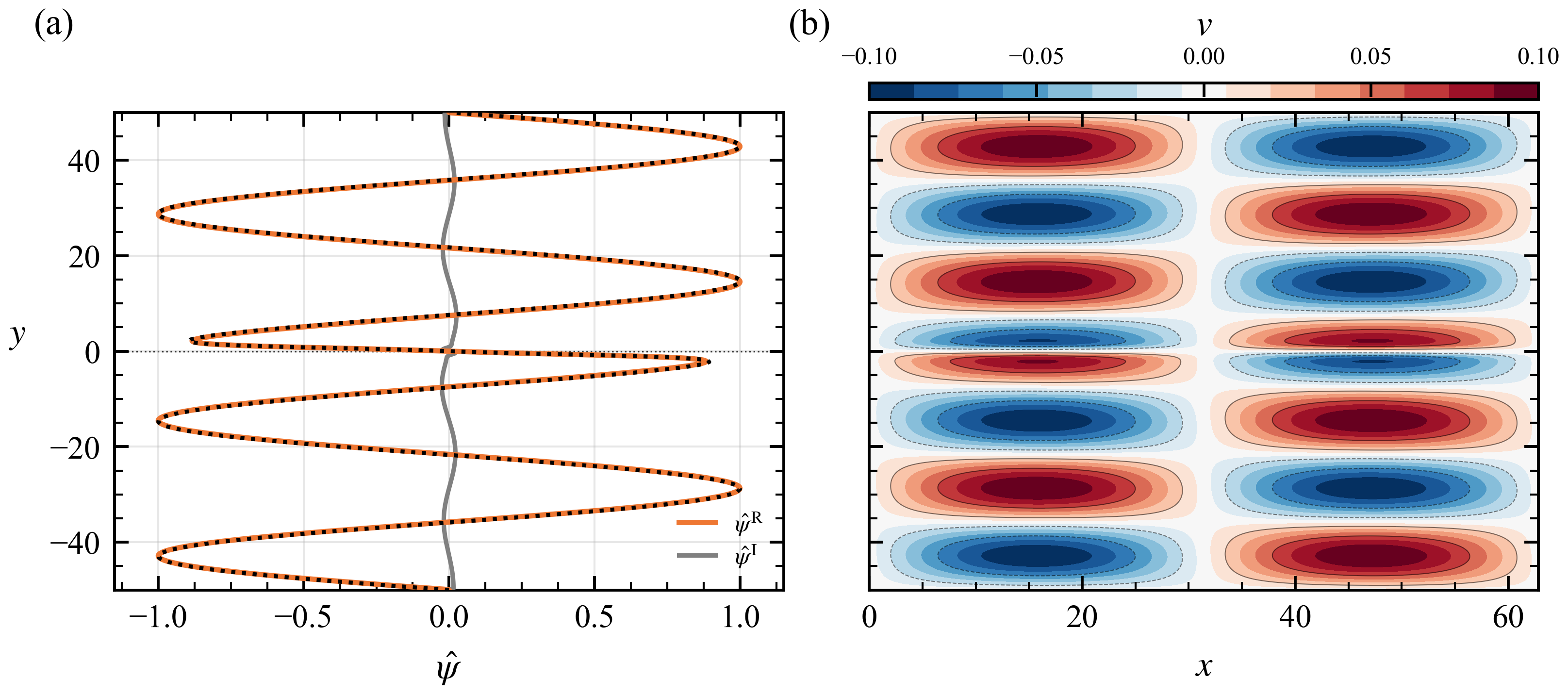}
    \caption{
    Eigenfunction of the unstable mode at $(k,J,H,Re,Pr)=(0.1,0.07,50,1000,1)$. (a) Real and imaginary parts of the Fourier coefficient. The black dotted curve is the solution of Taylor-Goldstein equation (\ref{q0eq}) with $c_0=1.09$. 
    %[Imaginary to grey, remove legend for TG][shall we use coherent notation for real/imag, e.g. $\hat{\psi}^R$ for real][dashed to dotted]
    (b) Vertical velocity distribution. %[panels are too big]
    %
%    fig 14: eigenfunction plot at the triangles and J=0.07 [plot imaginary on the same scale.]
    }
    \label{fig:varyingH_neutral_curves}
\end{figure}
In addition to the mode explored above, a variety of internal gravity wave structures emerge. %, as revealed by a more careful linear stability analysis. 
Figure 13a shows the neutral curves of the odd mode for $H=10,15,$ and 50, where the $H=10$ result is identical to that presented in figure 1a.
%At $k=0.1$, the critical value of $J$ does not vary monotonically as $H$ is increased, and at $H=50$ the neutral curve is made of piecewise smooth curve. 
The neutral curves change their shape in a rather complicated manner. 
This is because multiple distinct unstable modes appear and disappear as $H$ is varied; see panel (b) where the 
%top 8 
growth rates are plotted at $J=0.07, k=0.1$ (the circle at panel (a)).
Panel (d) is the associated phase speed. 
Figure 14 depicts the eigenfunction of the unstable mode obtained at $H=50$. Interestingly, the eigenfunction exhibits oscillations in the far field.

%The oscillations outside the jet represents the radiating internal gravity waves.

The behaviour of the stability results is explained by the viscous mode mechanism briefly discussed in section 3.1, where the leading order wave is governed by the neutral Taylor-Goldstein equation (\ref{q0eq}), while viscous effects enter at higher order. 
The black lines in figure 13d show the values of $c_0$ that give the eigenvalue $k^2=0.1^2$ in (\ref{q0eq}).
In figure 14a, the eigenfunction of the unstable branch (the black dotted curve) agrees very well with the numerical solution of (\ref{visstab}).
The Taylor-Goldstein equation (\ref{TGeq}) predicts radiation when the vertical wavenumber 
\begin{eqnarray}
\alpha(c)=\sqrt{\frac{J}{c^2}-k^2}\label{alphaeq}
\end{eqnarray}
is real-valued. 
The ratio of $H$ to the wavelength
$\lambda=2\pi/\alpha$ estimates the number of wave oscillations within the domain. 
%To estimate the number of oscillations within the domain, 
The variation of $H/\lambda$ along the neutral curve is plotted in figure 13c. 
The wavelength is relatively large, so clear radiation can only be observed for sufficiently large $H$.
However, excessively large $H$ stabilise the mode (see figure 13b), indicating a trade-off.
\begin{figure}
    \centering
    \includegraphics[width=0.9\linewidth]{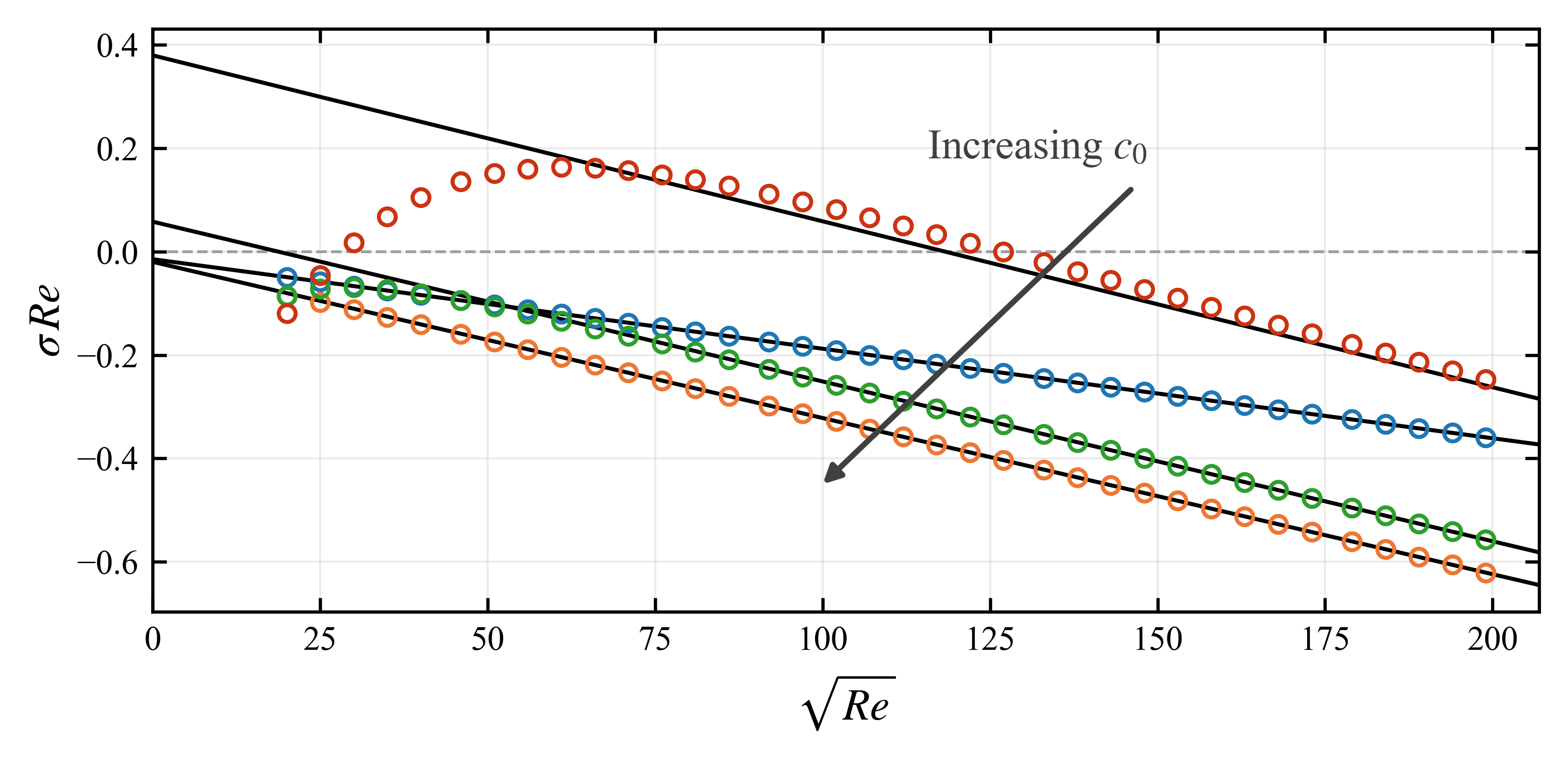}
    \caption{Large Reynolds number scaling of the growth rates at $(k,J,H)=(0.1,0.07,50)$. The symbols are the four largest growth rates found using (\ref{visstab}). The lines are the two-term asymptotic predictions from (\ref{twoterm}) for $c_0=1.0856,1.3324,1.7034,$ and $2.2603$. 
    }
    \label{fig:placeholder}
\end{figure}
% \Kengo{I have a no slip version that can do up to $\sqrt{Re}=400$, do you want $200$ to $400$ [I think the current ver is clear enough]}

Using the asymptotic expansion (\ref{expRe}) with $\epsilon=Re^{-1/2}$, 
the equations for $\psi_1$ and $\psi_2$ are obtained at $O(\epsilon)$ and $O(\epsilon^2)$, respectively:
\begin{subequations}
\begin{eqnarray}
\mathcal{L}_0\psi_1=c_1F_2\psi_0,\label{psi1eq}\\
\mathcal{L}_0\psi_2=c_2F_2\psi_0+c_1^2F_3\psi_0+c_1F_2\psi_1-i\mathcal{L}_v\psi_0, \label{psi2eq}
\end{eqnarray}
\end{subequations}
where %$F_2=\frac{\mathcal{U}\mathcal{U}''-2J}{\mathcal{U}^2}, F_3=\frac{\mathcal{U}\mathcal{U}''-3J}{\mathcal{U}^3}$.
\begin{eqnarray}
F_2=\frac{\mathcal{U}\mathcal{U}''-2J}{\mathcal{U}^2},\qquad F_3=\frac{\mathcal{U}\mathcal{U}''-3J}{\mathcal{U}^3},\\
\mathcal{L}_v\psi_0=k^{-1}\left \{\psi_0''''-2k^2\psi_0''+k^4\psi_0-\frac{J}{\mathcal{U}Pr}\left ( \left (\frac{\psi_0}{\mathcal{U}} \right)''-k^2\left (\frac{\psi_0}{\mathcal{U}}\right )\right )\right\}.
\end{eqnarray}
A standard technique for determining $c_1$ is to use the solvability condition, that is, to take the $L^2$ inner product 
\begin{eqnarray}
\langle f,g \rangle=\int^H_{-H}fg\,dy
\end{eqnarray}
of (\ref{psi1eq}) with the adjoint solution $\psi^{\dagger}$. An integral expression for $c_1$ is then obtained as
\begin{eqnarray}
c_1=\frac{B_1}{\langle \psi^{\dagger},F_2 \psi_0\rangle}.\label{c1eq}
\end{eqnarray}
From the Sturm-Liouville form (\ref{q0eq}), we can readily find $\psi^{\dagger}=\psi_0/\mathcal{U}$. The numerator 
%$B_1=-2m^{-1}(\psi_0|_{y=-H})^2$ 
\begin{eqnarray}
B_1=-\frac{(1+i)\{(\psi_0'|_{y=H})^2+(\psi_0'|_{y=-H})^2\}}{(2kc_0)^{1/2}}\label{B1B1}
\end{eqnarray}
can be found by the boundary term after integrating by parts, noting that $\psi_1$ is not zero at $y=\pm H$. 
The boundary value arises from the effects of the thin near-wall Stokes layer, as explained in detail in Appendix A. 
The same approach can be used to deduce
\begin{eqnarray}
c_2=\frac{B_2+i\langle \psi^{\dagger},\mathcal{L}_v\psi_0\rangle-\langle \psi^{\dagger} ,c_1^2F_3\psi_0+c_1F_2\psi_1\rangle}{\langle \psi^{\dagger},F_2 \psi_0\rangle}\label{c2eq}
\end{eqnarray}
from (\ref{psi2eq}), where $B_2$ is given by (\ref{B2B2}). 

The solid lines in figure 15 show the two-term asymptotic approximation of the scaled growth rate
\begin{eqnarray}
Re\,\sigma=Re^{1/2}kc_1^I+kc_2^I.\label{twoterm}
\end{eqnarray}
Note that $Re^{1/2}$ is used as the abscissa in the figure, so that (\ref{twoterm}) appears as a straight line. 
The points are the full numerical results continued from the 4 modes found at $H=50$ of figure 13b. 
We can show that $c_1^I$ is negative for sufficiently large $H$ (see Appendix A). Therefore, the second term is necessary to account for the instability, which should nevertheless disappear as $Re$ and $H$ become large. 

In addition to the odd radiating mode discussed here, an even radiating mode emerges at small $k$; see Appendix B.

\begin{figure}
    \centering
    \includegraphics[width=\linewidth]{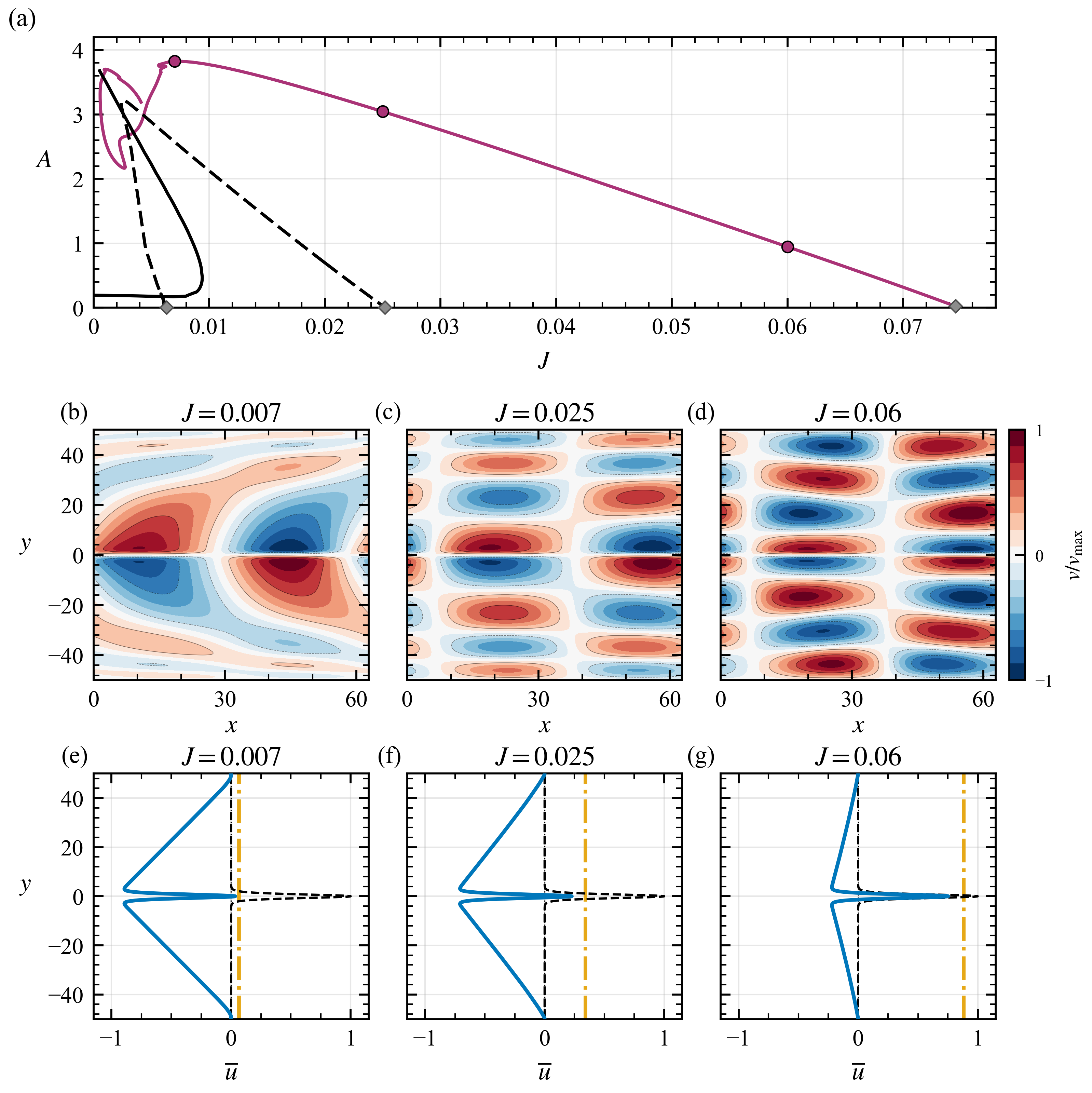}
    \caption{
    Bifurcation from the odd mode instability at $(H,k,Re,Pr)=(50,0.1,1000,1)$. (a) Bifurcation diagram. The purple curve shows the solution bifurcating from the most unstable mode. The black solid and dashed curves are the same as those shown in figure 10a. (b-d) Vertical velocity fields. The values of $v_\mathrm{max}=\max_{y,z}|v|$ are 0.01554, 0.02112, and 0.01512 for panels (b), (c), and (d), respectively. %at $J=0.007$, $J=0.025$, and $J=0.06$, respectively. 
    (e-g) Mean flows. See figure 5 for the format of the latter panels. 
    %The flow fields are plotted at the purple points indicated by the circles in panel (a). 
    %[write J in panels b-d]
    %Nonlinear results at $k=0.1$ for different branches at $H=50$
    % [can you try yellow dotted for c? If it looks good we may use the same format for other plots][values of J?]}
    % \Kengo{Might be a bit too bright ? I used gold yellow. I also attached the violet colour option [solid gold yellow? Violet seems too strong] [solid gold yellow looks pretty good, or maybe dashed gold yellow][Try dot dashed. always place yellow behind. Remove label c, we can hardly see][yellow to dot dashed, place this behind (same for fig 5, 8, 11)]
    }
 \label{fig:varyingJ_H50_k0p1}
\end{figure}

\subsection{Momentum transport by the radiating ECS}
Because of the mode switching, the bifurcation analysis from the neutral point at $(H,k,J)=(50,0.1,0.07455)$ shown in figure 13a does not yield the solution branches obtained in section 5.1. 
This is demonstrated in figure 16a, where %the bifurcation diagram for $(H,k)=(50,0.1)$ is replotted. 
the purple curve is the solution branch that bifurcates from the first eigenmode with the largest growth rate. 
The black curves are the same as those shown in figure 10a; of these, the dashed curve connects to the fourth and the fifth %\textcolor{black}{[Carl fix the sentence][I scanned through J at k=0.1 and found 5 neutral points, 0.07455, 0.0601, 0.0405, 0.025214, 0.00633]} 
eigenmodes at $J=0.025214$ and $0.00633$, respectively.

Along the purple branch in figure 16a, the vorticity field of selected solutions are shown in panels (b)-(d). 
%visualise the perturbation field by $v$. 
Shortly after the bifurcation (panel (d)), the radiated wave inherits the standing wave structure of the eigenfunction. This flow structure is, as expected, very similar to the nearby instability mode shown in figure 14b. 
As $J$ decreases, oblique patterns with opposite inclination angle above and below the jet gradually develop in the fluctuation field (panels (b) and (c)). 
For all cases in figure \textcolor{black}{16} we examined plots similar to those shown in figure 12 and confirmed that the far-field flow structures originate from internal gravity waves. 
%\Kengo{For all cases in figure 16? [fixed]}
Therefore, the same argument as in section 5.1 can be used to show that the mean flow in this region is approximately a linear profile (see figures 16e-g).

The waves seen in figure 16b locally resemble plane waves, in which case they are more theoretically tractable.
By writing $\tilde{\psi}$ as the sum of $e^{ik(x+\gamma y)}$ and its complex conjugate, we can show that the right hand side of (\ref{shearRS}) becomes $-2k^2\gamma$. 
According to panel (b), $\gamma$ is negative above the jet. In this case, (\ref{shearRS}) predicts positive mean shear, which is consistent with the observation in panel (e). 
The mean shear generates reverse flow, which decelerates the jet. As $J$ is reduced both the maximum velocity at $y=0$ and the phase speed $c$ decrease (panels (e-g)). In all cases, no critical level exists. 
%The positive phase speed $c$ in the plane wave approximation implies that the internal gravity waves in panel (b) propagate towards the jet in the reference frame considered here. 

We can also show that, for a perfectly standing wave structure, the right hand side of (\ref{shearRS}) is zero. This explains why the waves shown in figure 16d are relatively inefficient at generating mean shear. 

%However, by applying a Galilean transformation that shifts the minimum value of the mean flow to zero, it becomes clear that these waves originate from the jet.

\section{Strongly stratified regime}

\begin{figure}
    \centering
    \includegraphics[width=\linewidth]{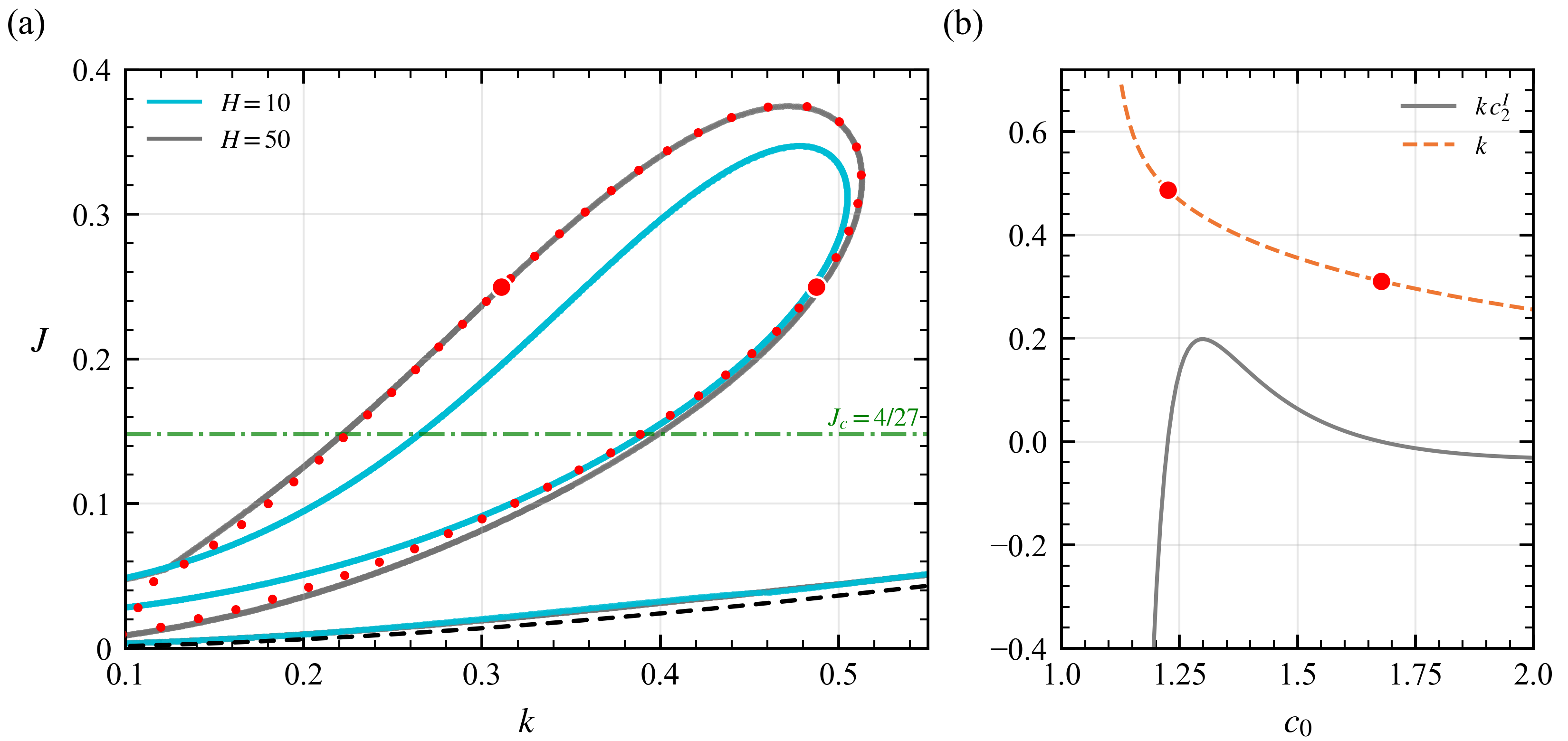}
    \caption{Stability analysis of the even mode. (a) Neutral curves in $k$--$J$ plane. The cyan solid and black dashed lines are the same as those shown in figure 1a. The grey solid line is the neutral curve obtained by increasing $H$ to 50 while keeping $Re=1000$. The red dots are the prediction from the asymptotic analysis.
(b) Intermediate step in the asymptotic analysis at $J=0.25$. %\textcolor{black}{$kc_2^I$ in panel b???} %[x axis $c_0$, blue to grey, $\sigma Re$ to $kc_1^I$ (or just $c_1^I$ if it looks good)[$kc_1^{I}$ looks better as diving by k would drag/stretch the grey curve even down more in $y$ axis.]]
    %[indicate two neutral points on the k curve][panel a needs simplification. Remove H=15. Use the same format as before for Drazin. Use purple for H=50. For asymptotic consider only retain selected points][in panel b use x range [1,2]?][Even mode, so H=10 to cyan.][Try H=50 to grey (put this to most behind).][Asymptotic blue dots to red dots]
    % [pink, grey Re=1000?] Yes. PCK20 paper noted this is an unresolved question on whether the region could shrink in the inviscid limit or the growth rate could vanish but the region remain a constant size. I used standard Chebyshev collocations sprectral element taking inspirations from my code a year or 2 ago. [Can you add more points for small J? I think now I can trust your code][added][H=50 and 70 are almost same?][yes they match quite well. data for $H=70$ I plotted mainly for $J >0.15$, and they are on the same curve with $H=50$.][I was thinking the right side has a slight mismatch, because they are on the lower side of $c>1$, and hence it might match if we increase $Re$, as the asymptotic solver has $O(1/Re^2)$ difference with finite $Re$.][at J=0.25 say, do you have k, c1 as function of c0. I guess you used them to plot panel b][yes, the curve in panel (b) would cross the $0$ threshold twice (most of the time), ans I'll use that phase speed value to trace back $k$.]
    }
    \label{fig:placeholder}
\end{figure}

In figure 1, the viscous even mode shows instability beyond the Miles-Howard stability criterion. 
Recall that the neutral curve was plotted using $H=10$.
At this parameter, the perturbation interacts with the walls (figure 3b). Two questions thus naturally arise: how the instability is influenced by the walls and whether it persists as $Re\rightarrow \infty$. 
In section 6.1, both questions are addressed at once by applying an analysis similar to that of \citet{parker2020viscous}.

Sections 6.2 and 6.3 analyse the travelling wave ECS bifurcating from the viscous even mode. DNS is performed and compared with the ECS. We use $H=10$ and $Re=1000$ for the DNS to make long-time computations feasible and to capture the interaction between the jet and the walls. We mainly consider strongly stratified flows for which inviscid instability is not possible. However, in section 6.3 we reduce $J$ to investigate the possible nonlinear interaction between viscous and inviscid modes.

\subsection{Asymptotic convergence of the neutral curve}
Figure 17 re-examines the neutral curve of the viscous even mode for $Re=1000$. The cyan curve shows the same result as that in figure 1a (i.e. $H=10$), while the grey curve is plotted for the larger domain height $H=50$. We confirmed that further increasing $H$ to 70 does not change the results from the latter case. 
Thus, we expect that the instability does not rely on the presence of the walls.

To confirm that this conclusion remains valid at arbitrary large $Re$, we perform the asymptotic analysis using the mapping
\begin{eqnarray}
z = \tanh\left(\frac{y}{L}\right).\label{maptanhL}
\end{eqnarray}
This map is a generalisation of 
(\ref{maptanh}), with the stretching factor $L>0$  introduced merely to improve numerical convergence.
We impose $\psi_0=0$ at $z=\pm 1$ to solve the Taylor-Goldstein equation (\ref{q0eq}) transformed into an ordinary differential equation in $z$, where the second derivative becomes
\begin{eqnarray}
\frac{d^2\hat{\psi}}{dy^2}=\frac{1}{L^2}\left ((1-z^2)^2\frac{d^2\hat{\psi}}{dz^2}-2z(1-z^2)\frac{d\hat{\psi}}{dz} \right).
\end{eqnarray}
The Chebyshev collocation method can be employed to determine $k$ as a function of $c_0$ (the red dashed curve in figure 17b).
The choice $L\approx 25$ gives optimal numerical convergence.

\begin{figure}
    \centering
    \includegraphics[width=\linewidth]{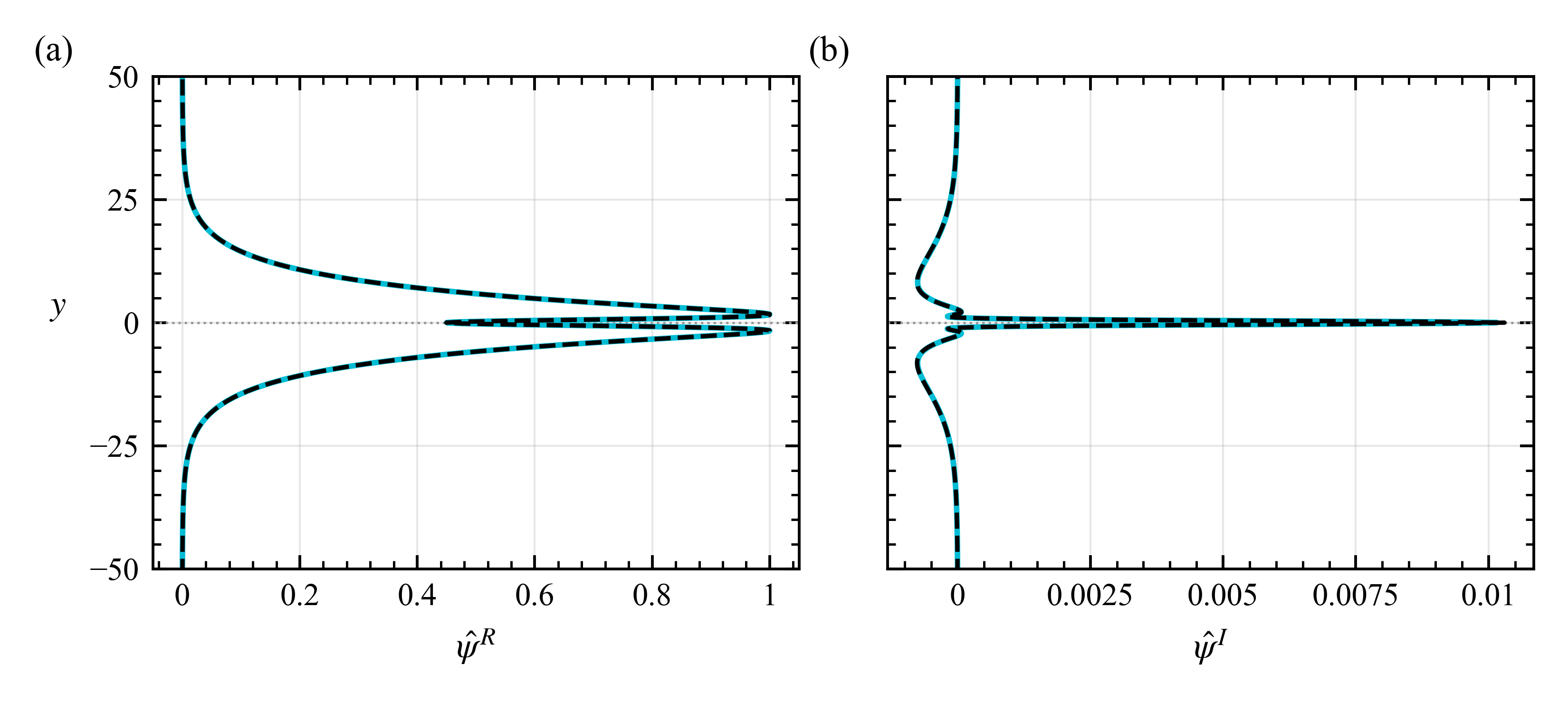}
    \caption{Eigenfunction of the unstable viscous even mode at $J=0.25, k=0.42$. The solid cyan curve is the numerical solution of (\ref{visstab}). The black dashed curve shows the approximation by the asymptotic analysis $\hat{\psi}=\psi_0+Re^{-1}\psi_2$. %[remove legends]
    %Comparisons of the eigenfuctions between the inviscid mode (part 1 solver to find $k$) and the viscous mode (normal Bickley solver to plot other figures in the main section) at $J=0.25$, $k \approx 0.4201$, $H=50$. The process was to fix $J$, and pick $c$, run the inviscid TG code to find $k$, then rerun the bounded viscous case with that $(k,J)$ and compare the eigenfunction on same scale $z$. $L=25$.[Actually this needs to be like Fig 2 ab as even mode][same brightness as fig 1, too dark]
    }
    \label{fig:placeholder}
\end{figure}
At higher order, we assume that the wall effects are negligible, so that $B_1=B_2=0$. In this setting, we obtain $c_1=0$ and 
\begin{eqnarray}
c_2=\frac{i\langle \psi^{\dagger},\mathcal{L}_v\psi_0\rangle}{\langle \psi^{\dagger},F_2 \psi_0\rangle},\label{c2Parker}
\end{eqnarray}
which is equivalent to the analysis in  \citet{parker2020viscous}, except for the use of the mapping.
Noting 
\begin{eqnarray}
\langle f,g \rangle=\int^{\infty}_{-\infty}fg\,dy=\int^1_{-1}fg\,\frac{L}{1-z^2}dz,
\end{eqnarray}
integration in (\ref{c2Parker}) can be performed numerically. The $c_0$ values of interest are greater than 1, so $\psi_0$ is purely real and hence $c_2$ is purely imaginary. The imaginary part \textcolor{black}{$c_2^I$ crosses zero twice, and so does the leading order growth rate $kc_2^I$} (see the grey curve in figure 17b), and the corresponding values of $k$ give the neutral curve (large red circles). 

Figure 17b is produced for $J=0.25$. The same computation can be repeated for various values of $J$, and the obtained neutral points are summarised by red circles in panel (a). 
The asymptotic results agree very well with the neutral curve for $(Re,H)=(1000,50)$.
Figure 18 shows the eigenfunction of the unstable mode obtained at $(J,k)=(0.25,0.42)$. The asymptotic approximation $\hat{\psi}=\psi_0+Re^{-1}\psi_2$ agrees excellently with the eigenfunction at $H=50$. Here, $\psi_2$ is obtained by solving the forced Taylor-Goldstein equation (\ref{psi2eq}) using the mapping (\ref{maptanhL}), subject to the boundary conditions $\psi_2=0$ at $z=\pm 1$. 
All the theoretical results presented above suggest that the walls have a negligible effect on the instability mechanism when $H$ is sufficiently large.

\subsection{Nonlinear results in the strongly stratified regime}

\begin{figure}
    \centering
    \includegraphics[width=\linewidth]{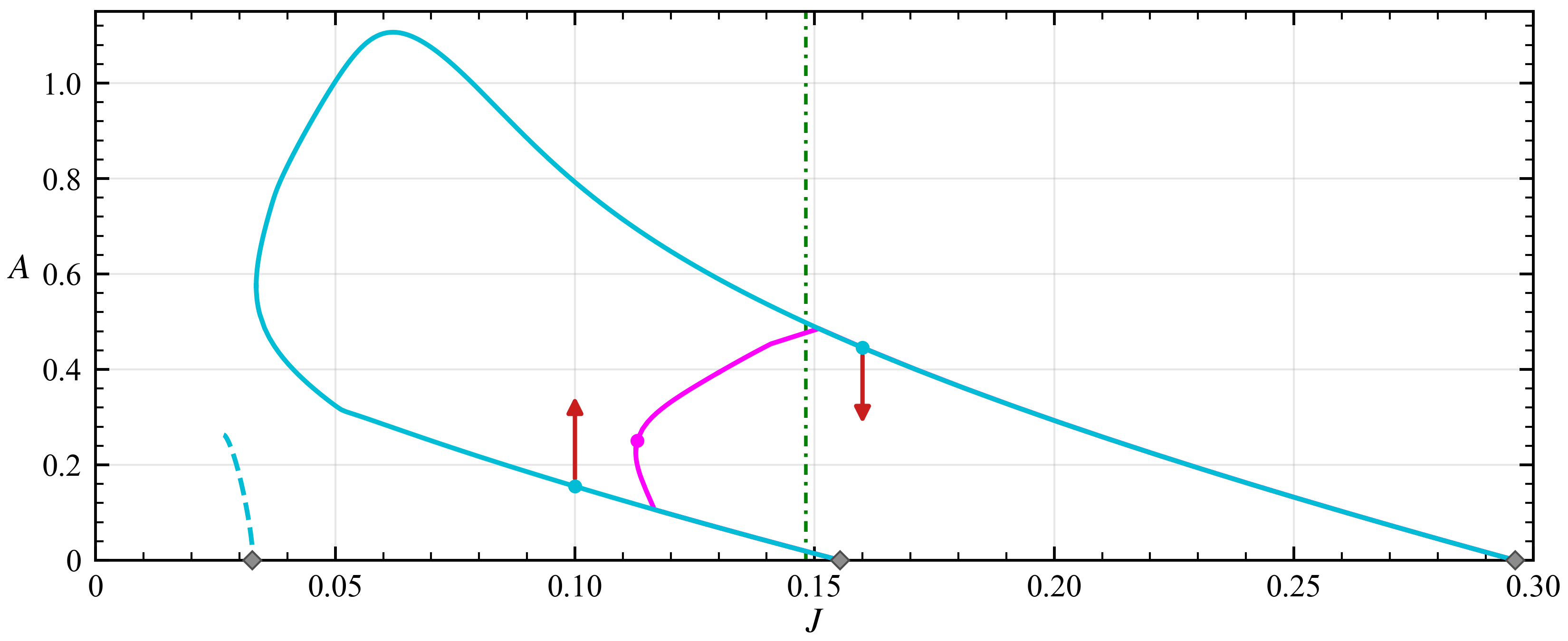}
    \caption{Travelling wave ECS bifurcating from the viscous even mode. Parameters are $(H,k,Re,Pr)=(10,0.4,1000,1)$. The cyan and magenta curves show the solutions with symmetric and asymmetric mean flows, respectively. 
   \textcolor{black}{The even viscous mode instability emerges between the two rightmost grey diamonds.}
    The dashed curve is the solution branch discussed in Appendix B. The red arrows indicate the DNS shown in figures 21 and 22. %[use magenta only for the bridged part][include green vertical dot-dashed]
    %[purple to cyan dashed][truncate as the appendix fig][same magenta as fig 7 (use same format as this fig)]
    }
    \label{fig:placeholder}
\end{figure}

\begin{figure}
    \centering
    \includegraphics[width=\linewidth]{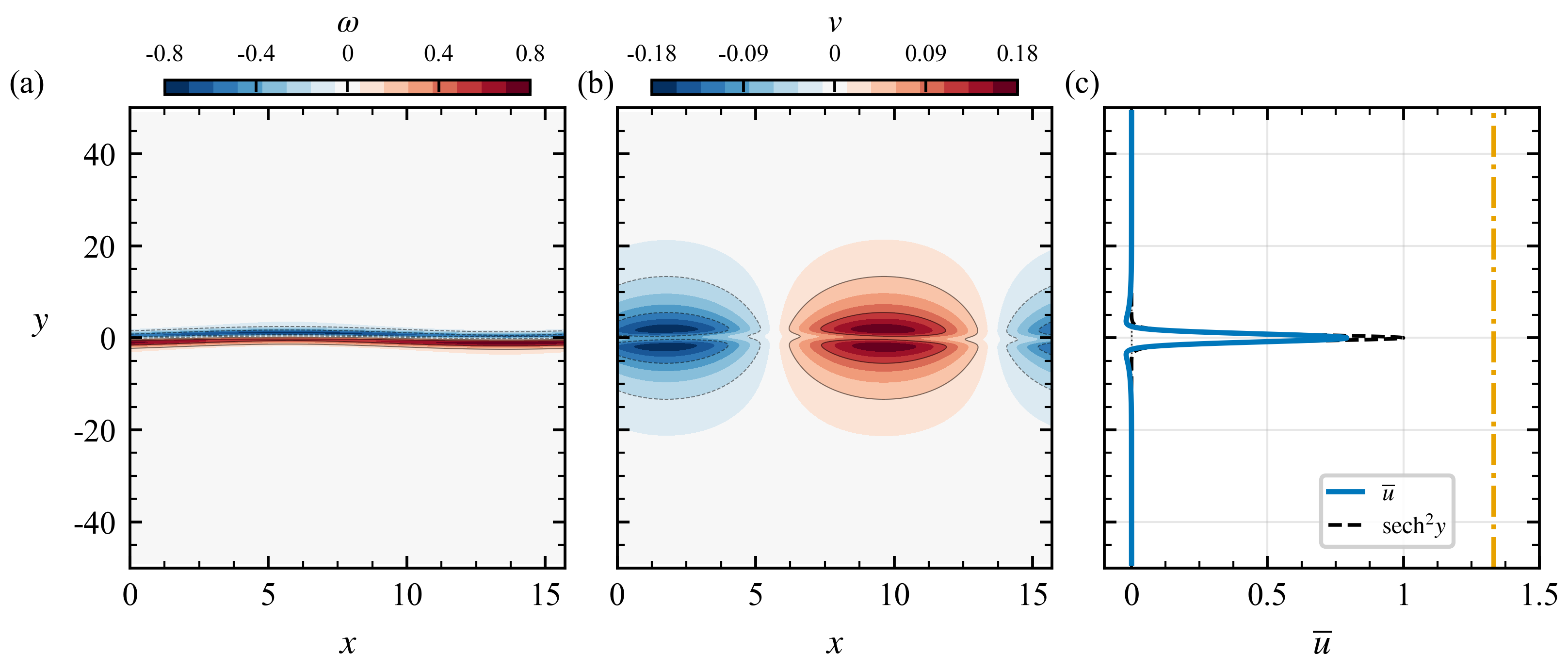}
    \caption{Travelling wave ECS obtained at $(k,J,H,Re,Pr)=(0.4,0.25,50,1000,1)$. Same format as figure 5.
    %$H=50$, $J=0.25$, $k=0.4$
    }
    \label{fig:placeholder}
\end{figure}

\begin{figure}
    \centering
    \includegraphics[width=\linewidth]{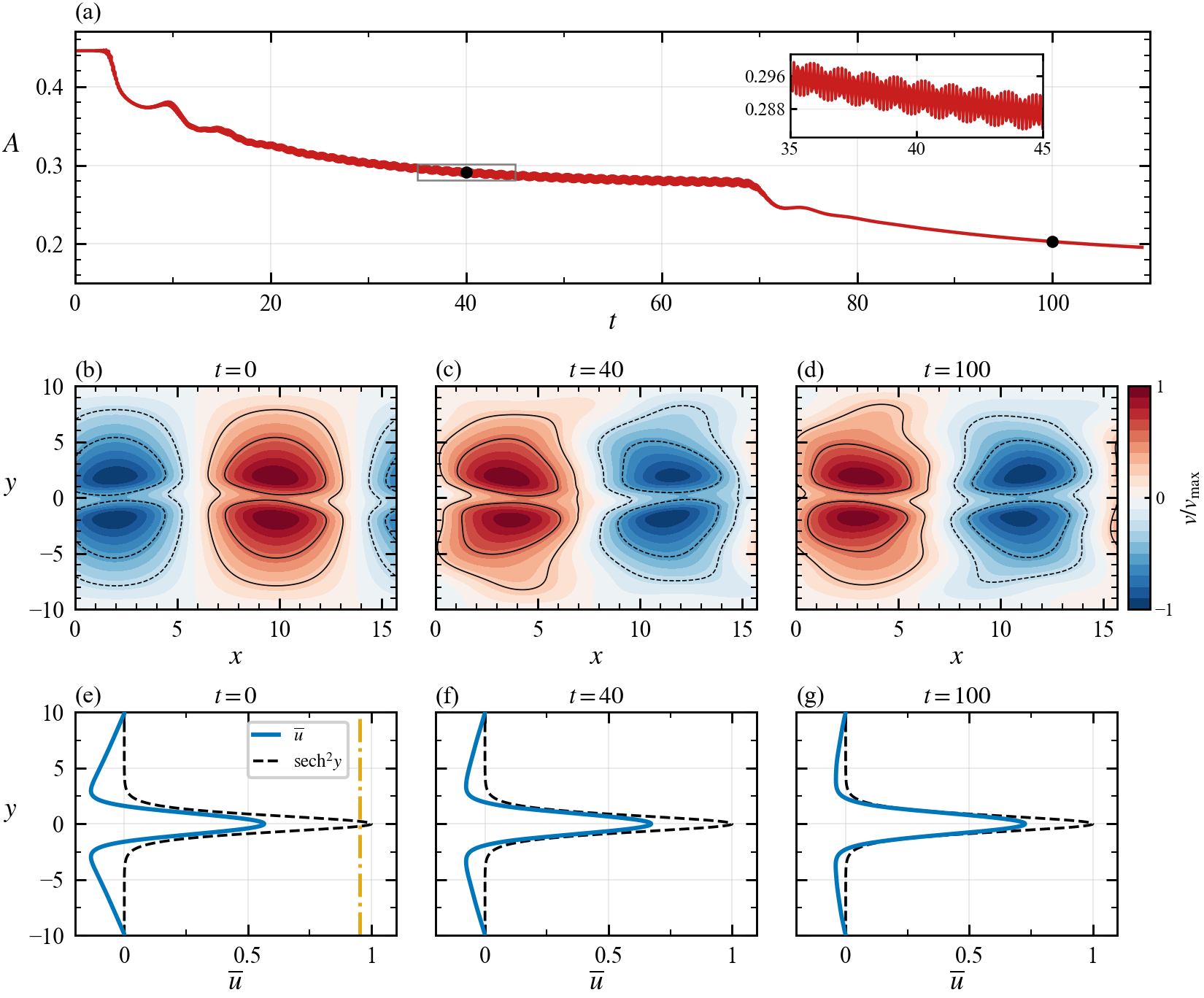}
    \caption{DNS results at $J=0.16$. Other parameters are $(H,k,Re,Pr)=(10,0.4,1000,1)$. 
    The initial condition is the ECS shown in figure 19. 
    (a) Time series of $A$. 
    (b-d) Vertical velocity fields. The values of $v_\mathrm{max}$ are 0.17081, 0.14386, and 0.14457 for panels (b), (c), and (d), respectively. 
    %at $t=0,10,$ and 40, respectively. 
    (e-g) Mean flows. The yellow vertical dot-dashed line is the phase speed $c$ of the ECS. %\textcolor{black}{[we may have to choose t=40, 100][place x tics to 10][do you have results up to 110? If so extend x axis][108.1 now, hopefully can get to 110 (you can put as far as you get before the seminar today, then extend 105 or 110)][also consider putting another inset around 100 if there are interesting oscillations]}%[add this][no only panel e, phase speed is undefined for f and g]
    %|v_\mathrm{max}|= 0.17081, 0.14386, 0.14457$ at $t=0$, $10$, and $40$ respectively.
    %[add mean flows]
    }
    \label{fig:placeholder}
\end{figure}
At $k=0.4$ there are three neutral points in figure 17, and ECS can be obtained from these points using the Newton method. The resulting bifurcation diagram is shown in figure 19. The solution branch bifurcating from the linear neutral point with $J=0.2962$ returns to the linear neutral point with $J=0.15548$.
%\textcolor{black}{$0.15548$}. 
This branch is the ECS associated with the even viscous mode discussed in section 6.1.  
There is another branch bifurcating from the even radiating mode at $J=0.03267$ (dashed curve) and is discussed in Appendix B.

We observed that small-amplitude solutions near the bifurcation point at $J=0.2962$ are insensitive to changes in $H$. 
For example, at $J=0.25$, a domain height of $H=50$ is sufficient to obtain coherent structures that are completely detached from the walls. As depicted in figure 20, their basic structure consists of sinusoidal meandering of the jet accompanied by large-scale vortices. The phase speed is below unity but remains faster than the mean flow.

The solution amplitude increases as $J$ decreases as shown in figure 19. 
Varying $H$ becomes difficult for $J\lesssim 0.15$, suggesting that the interaction between the coherent structures and the walls becomes increasingly prominent. 
%[correct?][that's right, H can go up to H=20 for J=0.15]. 
%
%, which corresponds to a value of $J$ slightly above the Howard-Miles criterion. 
Figure 21 shows the DNS result at $J=0.16$, initialised with the ECS at the starting point of the downward arrow in figure 19. 
Although the growth rate of the even viscous mode is small, its nonlinear development can produce a large distortion of the mean flow (panels (e)-(g)). 
The reverse flow in the mean flow is driven by non-radiating internal gravity waves generated outside the jet (panels (b)-(d)), through the mechanism described in section 5.1.

%The symmetry of the mean flow is well preserved for all $t$.

Small-amplitude oscillations in the time series of $A$ seem to arise from rapid fluctuations of the internal gravity wave.
%ECS is unstable to an oscillatory instability. 
%\Kengo{I think we need to be a bit careful here for $J=0.16$. [DNS still running?][yes DNS still running, it's dropped but starts to get stable again] I am not entirely sure if it has reach the statistically steady state yet I have a feeling it will drop again. $J=0.1$ I think is ok}
\textcolor{black}{Nevertheless, once the high-frequency signals of the oscillations are excluded, the flow evolves only slowly and remains qualitatively similar to the ECS throughout the long-time integration. The oscillation amplitude suddenly weakens around $t=70$, after which the reverse flow in the mean flow gradually diminishes.}

For $H=10$, the ECS is stable for $J\gtrsim 0.22$. 
DNS does not detect a transition from the travelling wave to another state.

\subsection{Nonlinear results in the moderately stratified regime}

\begin{figure}
    \centering
    \includegraphics[width=\linewidth]{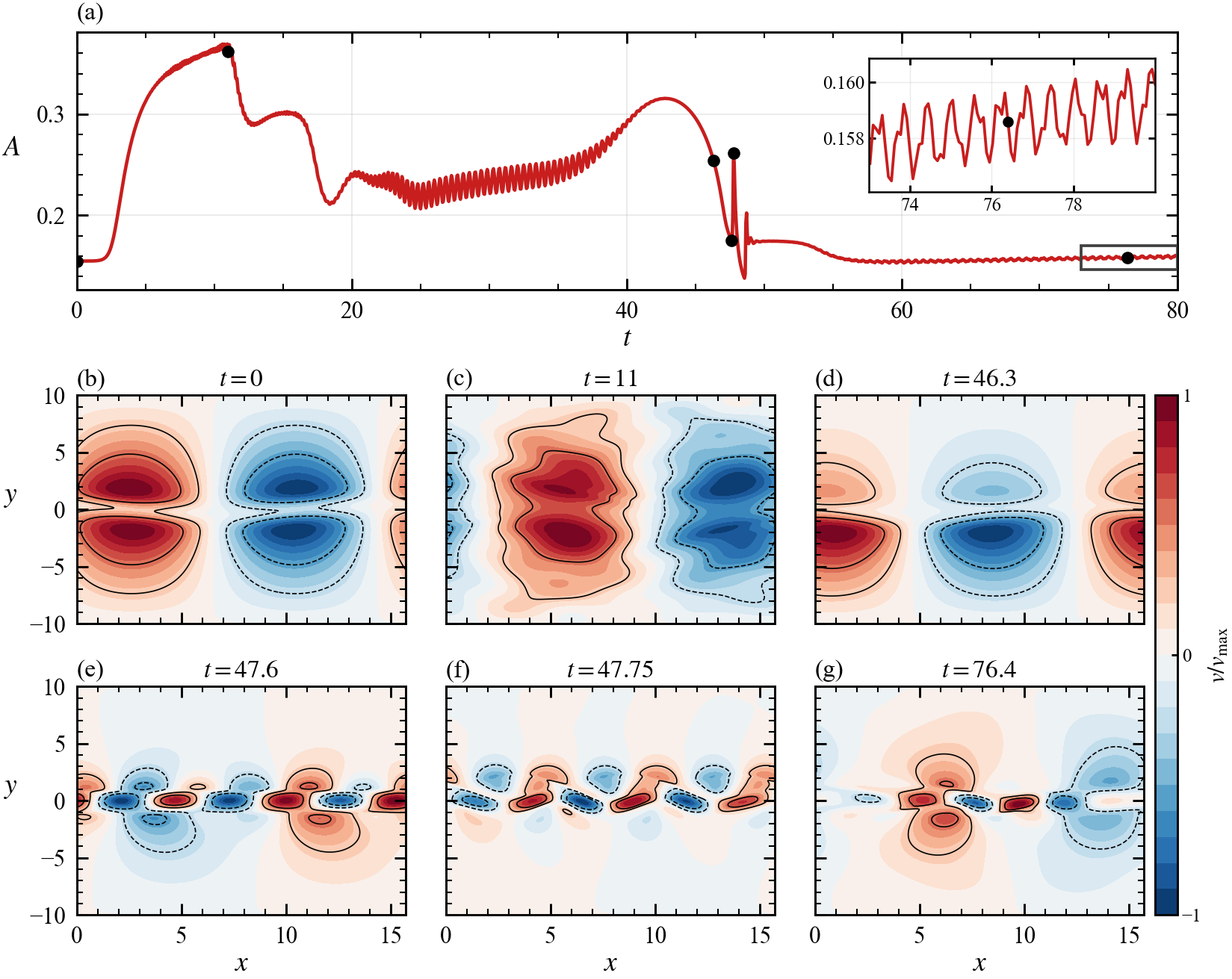}
    \caption{DNS results at $J=0.1$. Other parameters are the same as figure 21. The initial condition is the lower branch ECS in figure 19. 
    (a) Time series of $A$. (b-g) Vertical velocity fields. The values of $v_\mathrm{max}$ are
0.10745, 0.21662, 0.21173, 0.01481, 0.29568, and 0.12645 for panels (b) to (g), respectively.
    %$v_{max} = 0.10745,0.21662,0.21173,0.01481,0.29568,0.12645$ (from $b-g$, respectively) [vmax figs 16 and 21 as well]
    }
    \label{fig:placeholder}
\end{figure}

\begin{figure}
    \centering
    \includegraphics[width=\linewidth]{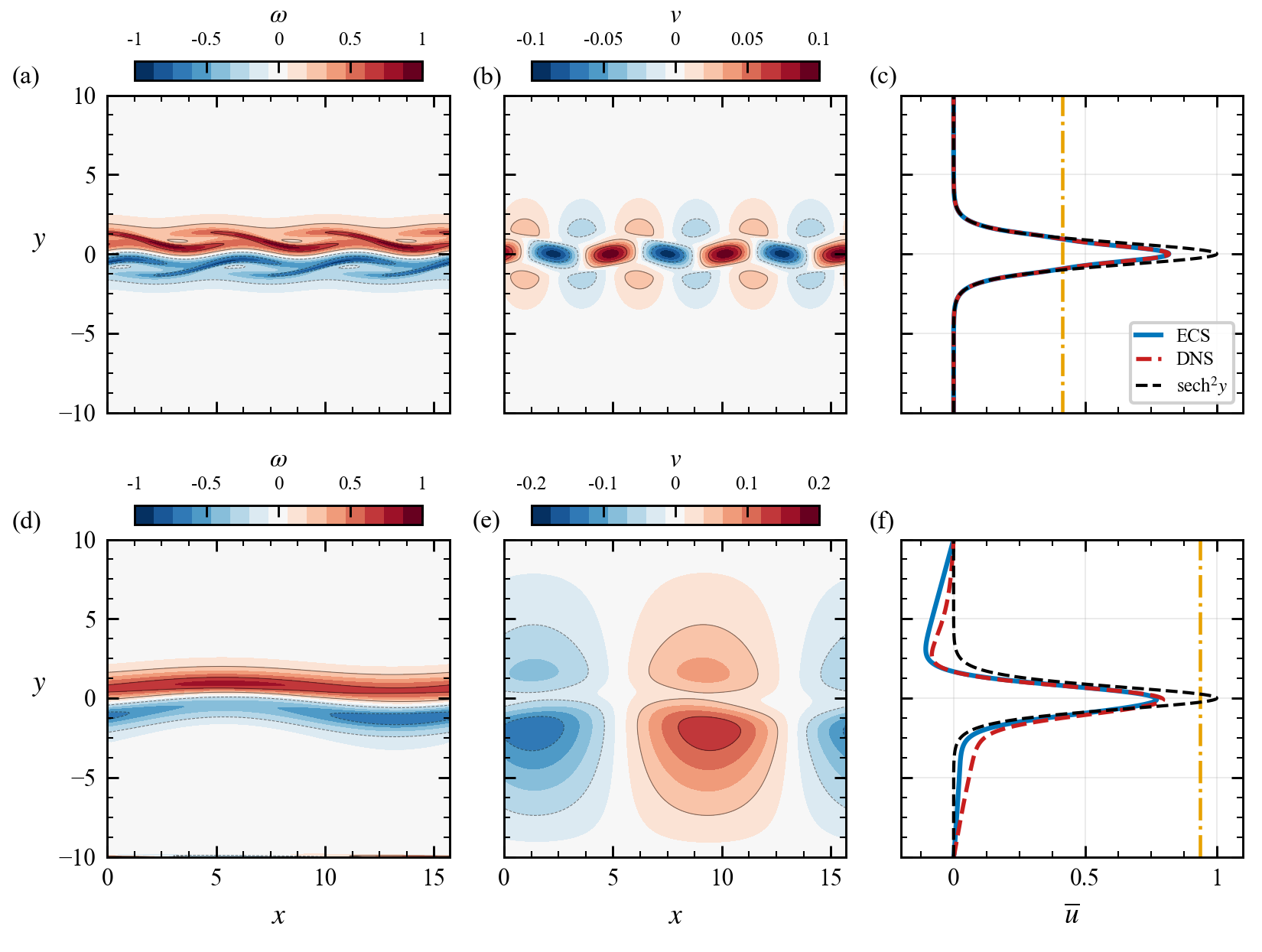}
    \caption{
Flow field of ECS relevant to the DNS in figure 22. The format is the same as figure 5 but the mean flow of a DNS snapshot is included (red dashed lines in panels (c) and (f)). The parameters $(H,Re,Pr)=(10,1000,1)$ are used, as in the DNS. (a-c) ECS bifurcating from the even inviscid mode at $k=1.2,J=0.1$. %The solution branch is continued from the neutral point $J=???$ to $J=0.1$. 
The DNS mean flow is taken at $t=76.4$. (d-f) Asymmetric ECS bifurcating from the even viscous mode ECS (magenta circle in figure 19). The DNS mean flow is taken at $t=46.3$.
    }
\end{figure}

We also performed DNS at $J=0.1$, using the same periodic domain in $x\in [0,2\pi/k]$ with $k=0.4$. At this value of $J$, the linear even viscous mode with this wavenumber is stabilised, but  the flow is instead subject to an inviscid linear instability. 
This is evident from figure 1a, where, at $J=0.1$, perturbation with $k=1.2$ lie within the unstable region (three copies of the perturbations fit within the periodic box).

There are two ECS branches associated with the even viscous mode (figure 19), and they may interact with the inviscid instability. 
The DNS shown in figure 22 is initialised with its lower branch, namely, the starting point of the upward arrow in the bifurcation diagram. The initial flow pattern in panel (b) is similar to those seen in figure 21. However, as expected from the time evolution of $A$ in panel (a), the subsequent dynamics  are very rich.

The first modulation to appear in the flow is the non-monotonic decay of the internal gravity wave towards the walls (panel (c)). 
This flow feature can be understood as the coexistence of radiating internal gravity waves with the large-scale vortices associated with the viscous even mode. 
The wave-like structure develops at the local maximum of $A$ near $t=11$ and persists until approximately $t=40$, when the oscillations of $A$ subside. Waves with a range of phase speeds and structures are generated outside the jet, giving rise to complex oscillations in the mean flow. The mechanism by which the waves feed back on the mean flow is similar to that discussed in section 5.1.
For $t\gtrsim 40$, an asymmetry gradually develops between above and below the jet (panel (d)). 
The fluctuations suddenly diminish around $t=47.6$ (panel (e), note that the value of $v_\mathrm{max}$ is very small), preceding the burst that follows. 
During the burst, 
three copies of the short-wavelength coherent structure emerge (panel (f)). 
A second burst then occurs, after which the flow field eventually settles into an almost statistically steady state  characterised by a \textcolor{black}{wave packet} (panel (g)). The final state appears to be weakly chaotic upon closer inspection (inset in panel (a)).

The nonlinear three-copy state observed in the DNS is most likely associated with the inviscid instability \textcolor{black}{briefly mentioned} at the beginning of this section. Therefore, we repeated the bifurcation analysis similar to that shown in figure 7 for $k=1.2$. 
Figure 23a-c shows the ECS continued from the linear neutral point $J=0.12252$ to $J=0.1$. 
The $v$ field is similar to that in figure 22f, except that the latter exhibits slight asymmetry. We confirmed that the ECS is stable in the short computational box $x\in [0,2\pi/1.2]$, \textcolor{black}{but it is unstable in the box used for the DNS. The detected subharmonic instability apparently} forms the wave packet observed in the final state of the DNS in figure 22g. 
The major role played by the ECS in the final DNS state is further evidenced by the excellent agreement between the mean flows compared in figure 23c.

As shown in figure 23d-f, ECS exhibiting an asymmetric structure similar to that in figure 22d can also be found. This ECS is obtained from the magenta branch in figure 19, which bifurcates from the even viscous mode ECS. 
Unfortunately, the asymmetric solution branch does not extend to $J=0.1$, so ECS found at $J=0.113$ is used in the figure. We speculate that this solution branch may undergo further bifurcations, producing ECS at $J=0.1$. However, the resulting ECS are unlikely to be travelling waves, and thus finding such solutions is \textcolor{black}{not pursued} in the present study. 

A short wavelength asymmetric solution similar to that seen in figure 8 also exists at $k=1.2$, but only for $J\lesssim 0.018852$. We therefore conclude that these solutions are irrelevant to the DNS in figure 22.

\section{Conclusion}

We have carried out linear stability and bifurcation analyses leading to ECS of a simple stratified jet over a wide range of parameters, including jet strength ($Re$), stratification strength ($J$), jet-wall distance ($H$), and wavenumber ($k$). For the various modes that emerge, we examine whether a well-defined large $H$ limit exists and, for moderate values of $H$, how the presence of the walls influences the results. At selected parameter values, DNS results are compared with the ECS.

In section 3, we classified the linear instabilities into the inviscid modes studied analytically by Drazin \& Howard (1966) and the newly identified viscous modes. The former is essentially a Kelvin-Helmholtz instability associated with the presence of a critical level. In contrast, the viscous modes do not involve a critical level and are not subject to the Miles-Howard criterion, similar to those found for the mixing layer by \citet{parker2019kelvin, parker2020viscous}.

%That is, they are classified as `bounded' according to the terminology of Drazin et al. (1979). 
Section 4 examines short wavelength regime, where the dynamics are expected to be dominated by inviscid sinuous (even) mode. 
At relatively small $J$, a travelling wave ECS with an asymmetric mean flow can be spontaneously generated in DNS. 
In general, short wavelength instabilities are trapped within the jet. 
Nevertheless, the asymmetry originates from a velocity displacement across the jet that does not diminish in the large $H$ limit. 

For long wavelengths, the waves emitted by the coherent structures in the jet attenuate slowly and interact more strongly with the walls. For sufficiently large $H$, some linearly unstable modes are accompanied by radiating internal gravity waves. In section 5.2, we analyse a varicose (odd) viscous radiating mode instability in detail, while Appendix B discusses an inviscid sinuous (even) radiating mode instability. These results provide the first evidence that radiating mode instabilities can arise from a Bickley jet even in a uniformly stratified fluid. 
The boundary conditions on the walls have a significant impact on the fate of the viscous radiating instability in the limit of large $Re$ and $H$. 
Near no-slip walls, a Stokes layer develops and stabilises the growth rate (section 5.2). However, with slip boundary conditions, the Stokes layer is eliminated, and the instability survives even in the limiting case (Appendix A). 
%Thus, a jet and its surrounding walls can interact even when they are far apart.

ECS associated with the long-wavelength instability generate fluctuations outside the jet through a nonlinear mechanism within the jet. These fluctuations propagate as internal gravity waves that approximately satisfy the Taylor-Goldstein equation. 
These waves transport momentum outside the jet, generating a reverse flow that in turn decelerates the jet centreline velocity (section 5.1). 
\textcolor{black}{This indirect mean-flow distortion mechanism contrasts with that seen in section 4, which is directly generated by the coherent structures in the jet. }

We also investigated the strongly stratified regime (section 6). The viscous sinuous (even) mode exhibits instability in a parameter range where the Miles-Howard criterion predicts the absence of inviscid instability. Asymptotic analysis conclusively demonstrated that the instability persists in the large $Re$, large $H$ limit. For sufficiently large $J$, the travelling wave ECS bifurcating from the viscous sinuous mode is either stable or forms the backbone of more complicated dynamics.
Reducing $J$ slightly below the Miles-Howard threshold, our DNS detect nonlinear interactions between the sinusoidal meandering originating from the viscous mode and a short-wavelength inviscid mode.

The above findings have implications for fluid intrusions discharged at their neutral-buoyancy level into a stratified ambient. The study of this problem dates back to the experiments by \citet{manins1976intrusion} and \citet{zuluaga1972flow},  and has recently been revisited using DNS by \citet{vu2026planar}. The focus of the DNS is mainly on the wide intrusion regime, where the bulk Richardson number $J$ is large ($J\sim(NW^2/Q)^2$ for a planar source of areal flux $Q$ and width $W$ in an ambient of buoyancy frequency $N$). 
\textcolor{black}{The streamwise velocity profile is fairly symmetric, with radiating waves of different propagation speeds emanating from the front tip of the intrusion. Among these waves, one particular wave overtakes the front tip of the intrusion; its propagation speed is faster than that of the intrusion, similar to the viscous modes. It was also noted that the waves in the DNS are strongly influenced by the domain height, unless the domain is sufficiently tall.  %(slip boundary conditions are used in the DNS). 
%
%The propagation speed of the wave is faster than that of the intrusion similar to the viscous modes, so that, for a sufficiently tall domain, the wave overtakes the front tip[there're lots of waves of different speeds, there's one specific waves that do overtake but does not seem to be controlled by ambient height AS LONG AS the ambient height is large enough]. 
Furthermore, the streamwise velocity recoeded in the DNS is distorted by the waves, creating reverse flow. The mechanism underlying this phenomenon might be explained by the theory presented in section 5. % in the far field[suggest removing far field], 
In general, whether internal gravity waves carry momentum and thereby decelerate the intrusion remains an open question, and our study may shed some light on this issue.}
%and radiating internal gravity waves are observed to be emitted from the intrusion, generating a reverse flow. 

DNS of the narrow intrusion regime (i.e. small $J$) was also conducted in the follow up paper \cite{vu2026jets}. In this regime, the laminar intrusion tends to break down into turbulence producing various coherent structures. In the early stages of breakdown, the centreline symmetry is lost due to the emergence of coherent structures reminiscent of the ECS associated with the inviscid asymmetric/symmetric sinuous mode. 
On the other hand, at the tip of intrusion, dipole vortices similar to those shown in figures 6b and 11 develop and detach from the main body. Radiating internal gravity waves are more prominent for small to moderate $J$ \textcolor{black}{than the wide intrusion regime.} 

The simple flow configuration studied in this paper is therefore useful, to some extent, for isolating the fundamental mechanisms responsible for local coherent structure generation in more realistic problems, such as intrusions. A better quantitative comparison between the ECS and the intrusion simulation would be possible by modifying the external forcing in the governing equations to reproduce the local velocity and density fields observed in the DNS. 
The sharp density gradients across the intrusion interface should promote Kelvin-Helmholtz instability and the generation of internal gravity waves.

Most geophysical problems involve slow spatio-temporal development, for which the standard ECS approach is difficult to apply. However, a method combining the computation of the slow-scale spatial evolution with local ECS calculations has recently been developed and applied to curved channel and boundary layer flows \citep{song2025spatial,song2026beyond}. 
%; \citet{song2026beyond}. 
This approach can be viewed as a generalisation of the framework proposed by \citet{deguchi2018free}, and can naturally be extended to stratified jets and mixing layers.
The method can be further extended to incorporate slow time dependence, which is necessary to predict the propagation speed of the intrusion. The results of the present study may also provide a useful foundation for such future studies.

\backsection[Acknowledgements]{The authors thank A.C. Slim for helpful discussions. This research was supported by Monash eResearch capabilities, including M3 and from the National Computational Infrastructure (NCI Australia), an NCRIS enabled capability supported by the Australian Government.}

\backsection[Funding]{This research was supported by the Australian Research Council Discovery Project DP230102188 and Australian Research Council Discovery Project DP220101660.} 

\backsection[Declaration of interests]{The authors report no conflicts of interest.}

\appendix
\section{Asymptotic analysis of the viscous mode}

\begin{figure}
    \centering
    \includegraphics[width=\linewidth]{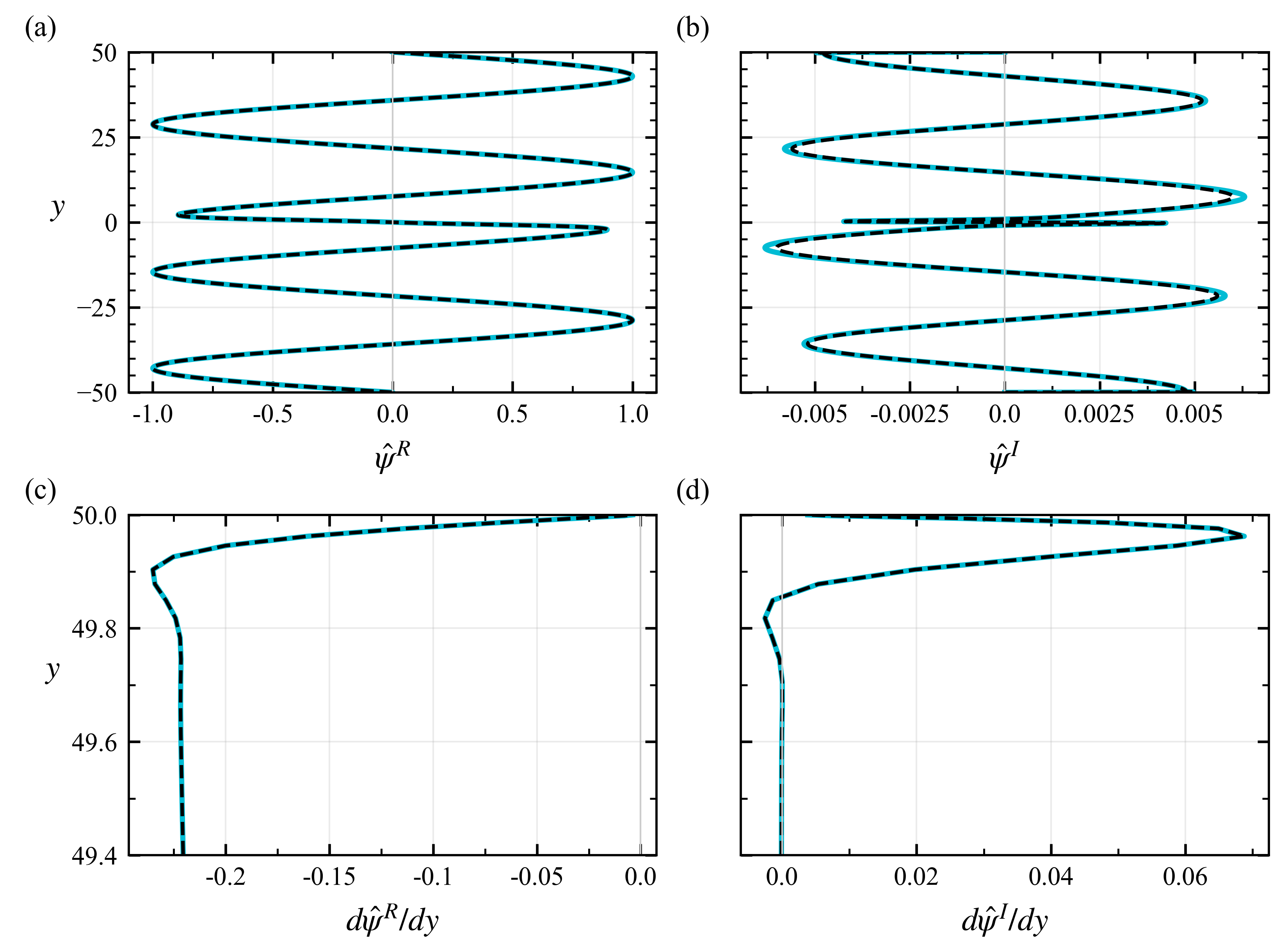}
    \caption{Validation of the asymptotic analysis at $(k,J,H)=(0.1,0.07,50)$. The cyan solid curves are the eigenfunction of (\ref{visstab}) with $(Re,Pr)=(10000,1)$. The black dashed curves are the composite solution (\ref{composite}). The solution with phase speed $c\approx 1.09$ is selected. The bottom two panels are the close up around the top Stokes layer. %[remove legends]
    }
    \label{fig:placeholder}
\end{figure}
The derivation of (\ref{c1eq}) and (\ref{c2eq}) necessitates an analysis of the Stokes layers. 
These layers have thickness $O(Re^{-1/2})$ and located near $y=\pm H$; so we introduce the stretched variable $Y=Re^{1/2}(y\mp H)$.
Asymptotic expansions in this region can be found as
\begin{subequations}\label{expStokes}
\begin{eqnarray}
\hat{\psi}=Re^{-1/2}\Psi_0^{\pm}(Y)+Re^{-1}\Psi_1^{\pm}(Y)+\cdots,\\ \hat{\rho}=Re^{-1/2}\Phi_0^{\pm}(Y)+Re^{-1}\Phi_1^{\pm}(Y)+\cdots.
\end{eqnarray}
\end{subequations}
Since $U$ is negligibly small at $y=\pm H$, substitution of (\ref{expStokes}) into (\ref{25eq1}) yields
\begin{subequations}
\begin{eqnarray}
-ikc_0\partial_Y^2\Psi_{j}^{\pm}=\partial_Y^4\Psi_j^{\pm},\\
-c_0\Phi_j^{\pm}+\Psi_j^{\pm}=\frac{1}{ikPr}\partial_Y^2\Phi_{j}^{\pm},
\end{eqnarray}
\end{subequations}
for $j=0,1$. 
The solution in the bottom Stokes layer is found as
\begin{eqnarray}
\Psi_j^{-}=\beta_j^{-}(Y+m^{-1}(e^{-mY}-1)), ~~~m=m_0(1-i),~~~m_0=(k c_0/2)^{1/2}.
\end{eqnarray}
It is easy to check that this solution satisfies the no-slip boundary conditions $\Psi_j=\partial_Y\Psi_j=0$ at $Y=0$ and 
$\partial_Y\Psi_j \rightarrow \beta_j^-$
as $Y\rightarrow \infty$. The matching conditions then yield
\begin{subequations}
\begin{eqnarray}
\psi'_0|_{y=-H}=\beta_0^-,\qquad \psi_1|_{y=-H}=-\beta_0^-/m,\\
\psi'_1|_{y=-H}=\beta_1^-,\qquad \psi_2|_{y=-H}=-\beta_1^-/m.
\end{eqnarray}
\end{subequations}
The top Stokes layer can likewise be analysed to obtain
\begin{eqnarray}
\Psi_j^{+}=\beta_j^{+}(Y-m^{-1}(e^{mY}-1))
\end{eqnarray}
and
\begin{subequations}
\begin{eqnarray}
\psi'_0|_{y=H}=\beta_0^+,\qquad \psi_1|_{y=H}=\beta_0^+/m,\\
\psi'_1|_{y=H}=\beta_1^+,\qquad \psi_2|_{y=H}=\beta_1^+/m.
\end{eqnarray}
\end{subequations}

The solvability conditions for (\ref{psi1eq}) and (\ref{psi2eq}) require us to work out the integration by parts
\begin{eqnarray}
\langle \psi^{\dagger},\mathcal{L}_0\psi_{j+1}\rangle=\left [\psi^{\dagger}\mathcal{U}^2(\frac{\psi_{j+1}}{\mathcal{U}})'-(\psi^{\dagger}\mathcal{U}^2)'(\frac{\psi_{j+1}}{\mathcal{U}})\right ]^H_{-H}\nonumber \\
=-m^{-1}(\beta_0^+\beta_j^++\beta_0^-\beta_j^-).
\end{eqnarray}
The right hand side is $B_{j+1}$.
For $j=0$, we obtain (\ref{B1B1}), while for $j=1$, we have
\begin{eqnarray}
B_2=-\frac{(1+i)\{\psi_0'|_{y=H}\,\psi_1'|_{y=H}+\psi_0'|_{y=-H}\,\psi_1'|_{y=-H}\}}{(2kc_0)^{1/2}}.\label{B2B2}
\end{eqnarray}
Having determined $c_1$ from (\ref{c1eq}), we can numerically solve (\ref{psi1eq}) for $\psi_1$. This provides all the ingredients needed to calculate $c_2$ from  (\ref{c2eq}), which in turn allows us to solve
(\ref{psi2eq}) for $\psi_2$. 
A composite solution can be found as
\begin{eqnarray}
\hat{\psi}=\{\psi_0+Re^{-1/2}\psi_1+Re^{-1}\psi_2\}+
\{Re^{-1/2}\Psi_0^-+Re^{-1}\Psi_1^-\}
+\{Re^{-1/2}\Psi_0^++Re^{-1}\Psi_1^+\}\nonumber \\
-\{\beta_0^-(y+H)-Re^{-1/2}(\beta_0^-m^{-1}-\beta_1^-(y+H))-Re^{-1}\beta_1^-m^{-1}\}\qquad\qquad\nonumber \\
-\{\beta_0^+(y-H)+Re^{-1/2}(\beta_0^+m^{-1}+\beta_1^+(y-H))+Re^{-1}\beta_1^+m^{-1}\},\qquad\qquad\label{composite}
\end{eqnarray}
%\Kengo{Is bottom wall $Y^{-} = Re^{1/2}(y+H)$. if so I think the sign needs to be swapped, as in $(y-H)$ becomes $(y+H)$, and vice versa [done, checked algebra?][Yes I think it looks ok just some typos earlier that's all.]}
where the last two terms are the common parts. The excellent agreement between this solution and the full numerical eigenfunction in figure 24 provides unequivocal validation of the asymptotic analysis.

\begin{figure}
    \centering
    \includegraphics[width=0.9\linewidth]{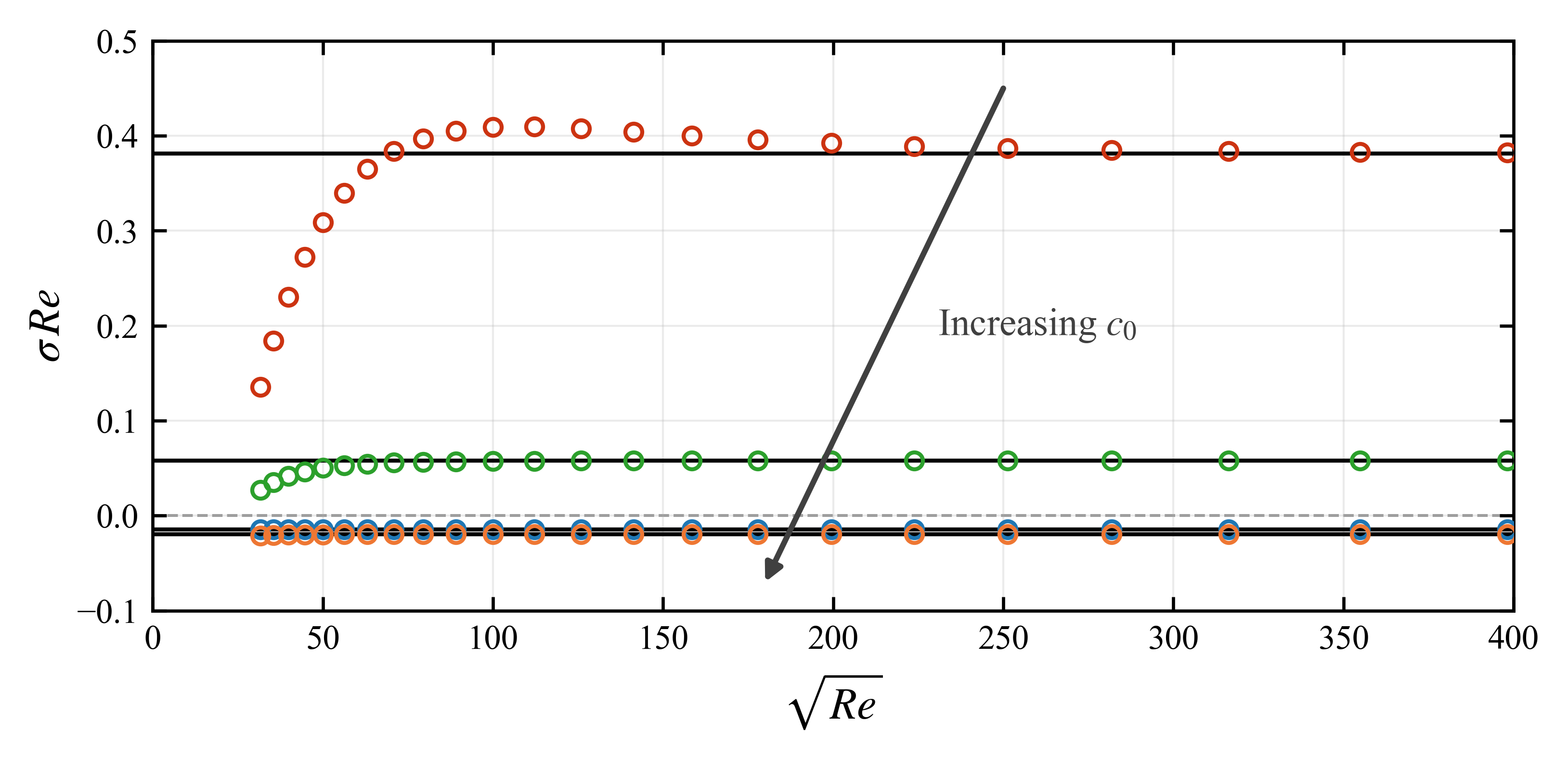}
    \caption{The same numerical results as in figure 15, but with slip boundary conditions imposed at the walls. The asymptotic predictions (solid lines) use $c_0=1.0856,1.3324,1.7034,$ and $2.2603$.
    %Caption [same format as fig 15][do we need arrows ?][yes and put c0 in the caption][y range to [-0.1,0.5]][you can use $\sqrt{Re}$ in horizontal]$c_0=1.0856,1.3324,1.7034,2.2603$ (similar to noslip)
    }
    \label{fig:placeholder}
\end{figure}
When $H$ is sufficiently large, $\psi_0$ behaves as $K\cos(\alpha(c_0) y+\phi)$ for large $y$, where $K$ and $\phi$ are  constants and  $\alpha(c_0)$ is the vertical wavenumber 
defined in (\ref{alphaeq}). 
This implies that if the Taylor-Goldstein equation (\ref{q0eq}) has a solution at $H=H_0(c_0)$ for some $c_0$, then it also has a solution at $H\approx H_0(c_0)+2n\pi/\alpha(c_0)$ for positive integers $n$ (this behaviour can indeed be confirmed in figure 13d). 
We now evaluate the integral in the denominator of (\ref{c1eq}) for the solution found at such $H$: 
\begin{eqnarray}
\langle \psi^{\dagger},F_2 \psi_0\rangle=\int_{-H}^{H}\frac{F_2}{\mathcal{U}}\psi_0^2dy=\int_{-H_0}^{H_0}\frac{F_2}{\mathcal{U}}\psi_0^2dy+2\int_{H_0}^{H_0+2n\pi/\alpha}\frac{F_2}{\mathcal{U}}\psi_0^2dy.
\end{eqnarray} 
Using the approximations $\psi_0\approx K\cos(\alpha y+\phi)$ and $\mathcal{U}\approx -c_0$ in the second integral, 
\begin{eqnarray}
\langle \psi^{\dagger},F_2 \psi_0\rangle\approx \int_{-H_0}^{H_0}\frac{F_2}{\mathcal{U}}\psi_0^2dy+\frac{4\pi J K^2}{\alpha c_0^3}n.
\end{eqnarray}
For sufficiently large $H$ (i.e. large $n$), the right hand side is positive, implying that 
$c^I_1<0$ in view of (\ref{c1eq}). Therefore, when the walls satisfy the no-slip boundary condition, the radiating modes must be stable when $H$ and $Re$ are large.

However, this conclusion does not carry over to the case of slip boundary conditions. The Stokes layer does not emerge, and the analysis becomes similar to that of \citet{parker2020viscous}. 
The Stokes layer contributions $B_1$ and $B_2$ vanish, as in section 6. Consequently, $c_1=0$, and the imaginary part of $c_2$ determines the asymptotic behaviour of the growth rate (see (\ref{twoterm})). 
Figure 25 shows results similar to those in figure 15, but with the boundary conditions replaced by slip conditions. 
The solid lines show the approximation (\ref{twoterm}), which is a constant due to the absence of the $c_1$ term. 
The figure clearly shows that the imaginary part $c_2^I$ can be positive for some modes, and the instability survives at large $Re$. 

%We can show slip results here. At $H=50$ you may do scaling analysis, plot $Re \sigma $ against $Re$. We recover the case of Parker.

\section{Even radiating mode}
We revisit figure 1, where the neutral curve is compared with the inviscid result from \citet{drazin1966hydrodynamic}. 
A closer look reveals that the even mode neutral curve deviates slightly from the analytic prediction for $k<0.7$. This deviation is in fact due to the even radiating mode. 

Figure 26a shows the bifurcation diagram obtained at $k=0.4$. The cyan dashed curve is computed with $H=10$ and thus corresponds to the result seen in figure 19. The grey dashed curve is the result computed at the higher domain height $H=50$. 

At the neutral point, the eigenfunction clearly exhibits radiation (panel (b)). From the eigenvalue the phase speed is computed as $c=0.213$; therefore, unlike the odd radiating mode discussed in section 5.2, critical layer singularities exist. This suggests that the mode is generated by an inviscid mechanism.
Panels (c-e) show the flow field of finite amplitude states at $J=0.03$. 
Radiating waves can also be seen in panel (c), and they produce a slight reverse flow in the mean flow (panel (e)).

\begin{figure}
    \centering
    \includegraphics[width=\linewidth]{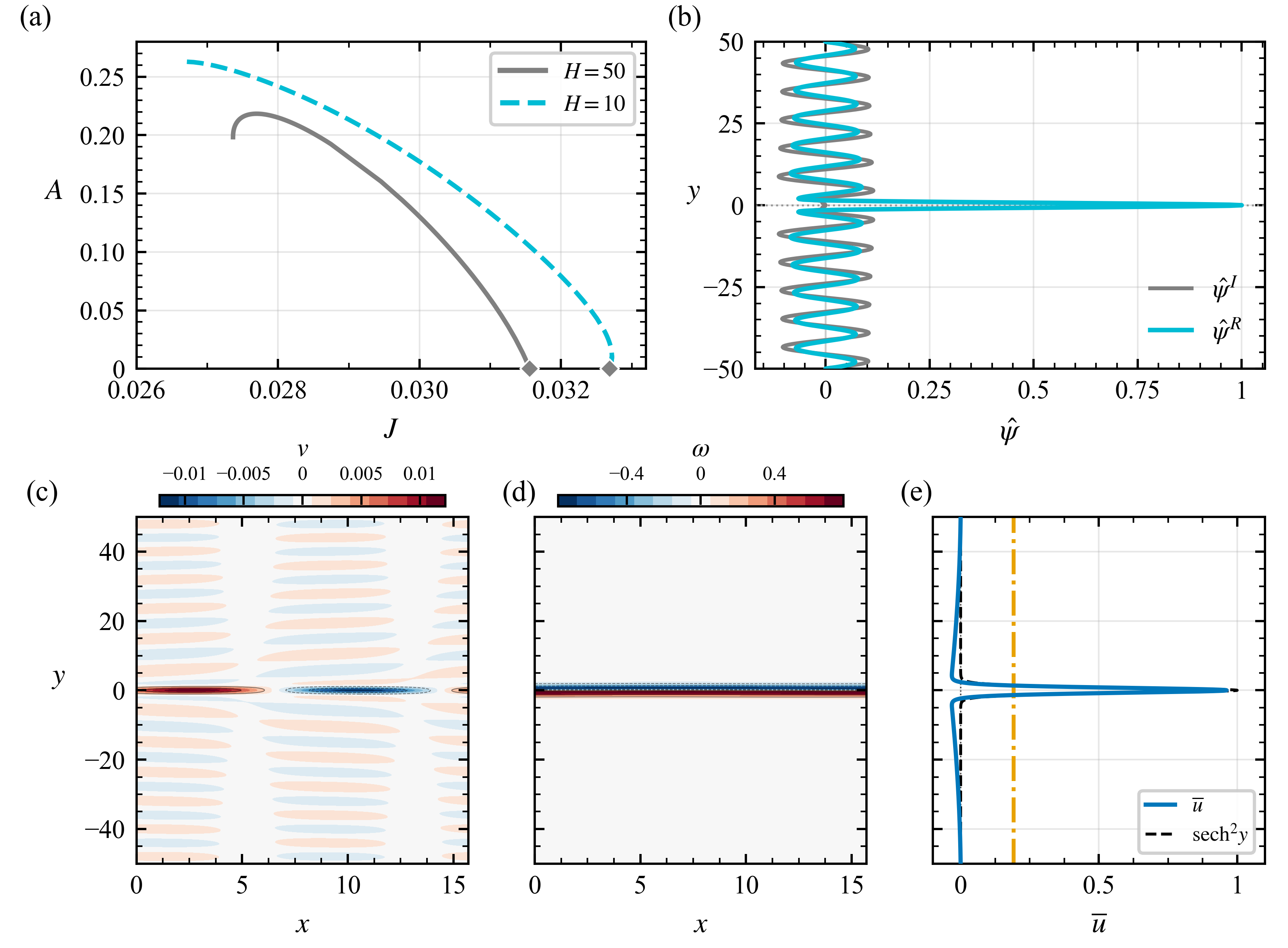}
    \caption{Bifurcation of the radiating even mode at $(k,Re,Pr)=(0.4,1000,1)$. (a) Bifurcation diagram. (b) Eigenfunction at the neutral point ($J=0.03155$) for $H=50$. (c-e) Flow field of the ECS at $(H,J)=(50,0.03)$. The same format as figure 5. 
    %[combine with the next figure] 
    %
%    $k=0.4$, $J=0.03155$, $c=0.213$, $H=50$ [viscous?]. (a) Bifurcation analysis from bifurcation point (nonlinear solution). (b) Viscous bounded solver at the same parameters in (a) (using LnBickley.for code to plot evec.txt)    
%    [left panel is neutral linear solution? Why the plot has been changed][earlier plot was the inviscid code I was testing. This new one is the viscous standard code][just plot solution without approximation, H=50 and Re=1000][to confirm, panel (b) you wanted the nonlinear solution or linear solution at the neutral point? I am plotting linear one.][linear solution at the neutral point][ok then the figure is what it is at the moment.][panel a, H=10 to cyan dashed]
    }
    \label{fig:placeholder}
\end{figure}
% \Kengo{Figure 26 I am not sure what's the idea of the right side panel. Do you want the viscous bounded eigenfunction, or the inviscid bounded eigenfunction? [only viscous. I guess you plotted viscous?][no I think I was plotting inviscid (can't remember if the code is even correct there). Viscous looks different, it's not attenuating ? Does that mean we can also do some analysis there? oh but it has critical layer which made it tricky...]}

%\newpage
\bibliography{references}
\bibliographystyle{jfm} 
\end{document}